\documentclass[submission,copyright,creativecommons]{eptcs}
\providecommand{\event}{FMAS 2025} % Name of the event you are submitting to
\usepackage[table]{xcolor}
\usepackage{iftex}
\usepackage{longtable}

\ifpdf
  \usepackage{underscore}         % Only needed if you use pdflatex.
  \usepackage[T1]{fontenc}        % Recommended with pdflatex
\else
  \usepackage{breakurl}           % Not needed if you use pdflatex only.
\fi

\usepackage{fretish}
\usepackage{paralist}
\usepackage{fretish}
\usepackage{graphicx}

\title{Towards A Catalogue of Requirement Patterns for Space Robotic Missions}
\author{Mahdi Etumi \qquad\qquad
Hazel M. Taylor \qquad\qquad Marie Farrell
\institute{University of Manchester\\
Manchester, UK}
\email{\quad\qquad firstname.surname@manchester.ac.uk \quad\qquad}
}
\def\titlerunning{Towards A Catalogue of Requirement Patterns for Space Robotic Missions}
\def\authorrunning{M. Etumi, H. M. Taylor \& M. Farrell}
\begin{document}
\maketitle

\begin{abstract}
In the development of safety and mission-critical systems, including autonomous space robotic missions, complex behaviour is captured during the requirements elicitation phase. Requirements are typically expressed using natural language which is ambiguous and not amenable to formal verification methods that can provide robust guarantees of system behaviour. To support the definition of formal requirements, specification patterns provide reusable, logic-based templates. A suite of robotic specification patterns, along with their formalisation in NASA's Formal Requirements Elicitation Tool (FRET) already exists. These pre-existing requirement patterns are domain agnostic and, in this paper we explore their applicability for space missions. To achieve this we carried out a literature review of existing space missions and formalised their requirements using FRET, contributing a corpus of space mission requirements. We categorised these requirements using pre-existing specification patterns which demonstrated their applicability in space missions. However, not all of the requirements that we formalised corresponded to an existing pattern so we have contributed 5 new requirement specification patterns as well as several variants of the existing and new patterns. We also conducted an expert evaluation of the new patterns, highlighting their benefits and limitations. %Based on these results, we identify future research directions including extensions to FRET. 
\end{abstract}

\section{Introduction}
\label{sec:intro}
Autonomous space robotics are used and planned to be used in a wide variety of missions. Examples include the ISS-based Astrobee (\cite{Bualat2018Astrobee}), the roaming Mars rovers (\cite{Mars2020MissionOverview}), and isolated satellites (\cite{AuthorCubeSatFlightSoftwareFramework}, \cite{Author2015MethodologyComplexSpaceMissions}). These missions help to further our understanding of the universe as well as enable Critical National Infrastructure to function correctly (e.g. communications, navigation systems, etc.). As a result, space robotic missions are viewed as both safety and mission-critical endeavours which require high degrees of assurance prior to and during deployment. Such complex critical systems typically necessitate a robust development process, often involving formal methods, that begins with a requirements engineering phase where developers elicit and formalise properties that capture correct system behaviour \cite{Engineeringautonomy}. 

Alongside requirements engineering, system specifications are constructed to enable the analysis of system architectures and behaviour prior to deployment. Requirements are typically expressed as natural language descriptions but formal specification frameworks use formal logics (rather than natural language) to specify and subsequently verify system behaviour. Various tools exist to help bridge the gap between natural language and logical requirements. One notable example is NASA's Formal Requirements Elicitation Tool (FRET) \cite{giannakopoulou2020formal}, which provides a structured natural language for requirement definition. From this structured natural language, called \fretish{}, FRET automatically generates a formal, temporal logic semantics for each requirement that can be used by formal methods.

It has long been recognised that specification is a huge bottleneck in critical systems development, especially when formal methods are used for reliable verification in the development of autonomous systems \cite{rozier2016specification}. Deriving usable specification patterns can provide developers with guidance during these initial project phases and a corpus of specification patterns for robotic missions already exists in \cite{Menghi2019SpecPatternsRoboticMissions} which was further extended by \cite{Vazquez2024RoboticMissionFRET}. These specification patterns are expressed in Linear Temporal Logic (LTL)  \cite{Menghi2019SpecPatternsRoboticMissions} and formalised using FRET \cite{Vazquez2024RoboticMissionFRET}. These existing catalogues of specification patterns are domain-agnostic in nature, which is undoubtedly a strength of the approach. However, it remains to be seen how these patterns can be used in a domain-specific context. In this paper, we recount our explorations about how these pre-existing patterns can be used in space systems. To that end, we contribute:\smallskip

\begin{compactitem}
    \item[1.] A corpus of requirements for space missions obtained through a detailed literature review. \smallskip
    \item[2.] An expanded catalogue of space domain-specific specification patterns that extends \cite{Menghi2019SpecPatternsRoboticMissions,Vazquez2024RoboticMissionFRET}, including both the LTL and FRET formalisations for each of the newly defined patterns. \smallskip
    \item[3.] An expert evaluation where we interviewed several experts with experience in the space domain to provide additional insights into our new patterns and potential future extensions.  \smallskip
\end{compactitem}

The remainder of this paper is structured as follows. We begin by providing an overview of the essential background material (Section \ref{sec:related}). Then, Section \ref{sec:methodology} describes the methodology that we used for our literature review for this study. This is followed by a detailed analysis of our findings in Section \ref{sec:requirements}. Here, we discuss the requirements that we found, novel patterns, metrics and an expert evaluation. We provide a detailed discussion in Section \ref{sec:discussion} and Section \ref{sec:conclude} concludes while outlining future research directions.

\section{Background}
\label{sec:related}

Specification patterns have been proposed in the literature as a way to bridge the gap between natural language descriptions and formal, logical patterns \cite{dwyer1999patterns}. The ultimate goal is to build a set of useful and usable patterns that can guide engineers during the requirements elicitation and formalisation steps. Specification patterns have been introduced for safety  \cite{dwyer1999patterns}, probabilistic \cite{grunske2008specification} and real-time specifications \cite{konrad2005real}. A comprehensive overview of various patterns has been conducted in \cite{autili2015aligning}. These specification patterns have been studied in various areas, including web services \cite{bianculli2012specification} and robotics \cite{Menghi2019SpecPatternsRoboticMissions,Vazquez2024RoboticMissionFRET,menghi2022mission}. 

All of these patterns provide template requirement specifications that can be used to formally specify suitable properties. In fact, they can be used as a guide for engineers who are not familiar with formal specification to elicit and formalise formally verifiable requirements. These sets of requirement specification patterns generally define generic, domain-agnostic requirement specifications. In this paper, we explore the use of existing (domain agnostic) specification patterns in space applications and identify new patterns from the space domain.

The core related work for this paper is the existing pattern catalogue for robotics in \cite{Menghi2019SpecPatternsRoboticMissions,Vazquez2024RoboticMissionFRET}. In these papers, Linear Time Temporal Logic (LTL) \cite{fisher2011introduction} is the logic of choice for robotic specification patterns. This is a sensible choice since multiple surveys on formal specification and verification for autonomous robots demonstrated that LTL is the predominant logic used in specifying requirements for  robotic systems \cite{luckcuck2019formal,azaiez2025revisiting}. As such, we also specify our patterns in LTL. LTL formulas are made up of atomic propositions, logical connectives and temporal operators. We provide the syntax of LTL:

\centerline{$\phi := \ true \ | \ a \ | \ \lnot \phi \ | \ \phi_1 \land \phi_2 \ | \ X \phi \ |  \ F \phi \ | \ G \phi \ | \ \phi_1 \mathcal{U} \phi_2 \ $}

\noindent where $a$ denotes an atomic proposition, $\land$ denotes logical conjunction and $\lnot$ denotes logical negation. Then $X \phi$ specifies that $\phi$ holds in the \textit{next} state, $F \phi$ that $\phi$ holds  \textit{eventually} in some future state and $G \phi$ that $\phi$ holds \textit{globally} in all future states. The \textit{until} operator, $\phi_1 \mathcal{U} \phi_2$, specifies that $\phi_1$ holds true until a future state where $\phi_2$ becomes true, and $\phi_2$ does indeed eventually become true. This is commonly known as \textit{strong} until in the literature. FRET (described below) uses the \textit{weak} version of until which omits the condition that $\phi_2$ must eventually become true. We use $\phi_1 \mathcal{W} \phi_2$ to denote weak until.

Requirements are typically specified using natural language, which is prone to ambiguity. For formal verification, formal logics are used to specify requirements. Various tools and approaches, including NASA's Formal Requirements Elicitation Tool (FRET) \cite{giannakopoulou2020formal}, bridge the gap between natural language and formalised, logical requirements. FRET uses a structured natural language, called \fretish{}, to give a clear syntax that users specify their requirements in. For each \fretish{} requirement, FRET generates a corresponding temporal logic semantics \cite{giannakopoulou2021automated}. The \fretish{} fields are \textcolor{Mahogany}{\small{\texttt{scope}}}; \textcolor{orange}{\small{\texttt{condition}}}; \textcolor{ForestGreen}{\small{\texttt{component*}}}; \texttt{\small{shall*}}; \textcolor{cyan}{\small{\texttt{probability}}}; \textcolor{blue}{\small{\texttt{timing}}}; and \textcolor{violet}{\small{\texttt{response*}}}. Fields marked with an asterisk (*) are mandatory. 

The \textcolor{Mahogany}{\small{\texttt{scope}}} field specifies the \emph{mode of operation} that is relevant to the component's behaviour. The  \textcolor{orange}{\small{\texttt{condition}}} field specifies the \textit{conditions} under which the requirement should hold. The \textcolor{ForestGreen}{\small{\texttt{component}}} field identifies the \textit{component} that the requirement applies to. The \textcolor{cyan}{\small{\texttt{probability}}} field specifies the \textit{probability} of the timed response, while the  \textcolor{blue}{\small{\texttt{timing}}} field specifies when the boolean \textcolor{violet}{\small{\texttt{response}}} should hold. For each field, there are various options for the keywords that can be used; these are designed to be intuitive to the user. For example, we can write requirements like:

\centerline{\scope{in FailSafe mode} \conditionF{if errorObserved} \component{Robot} shall \probability{with probability $\geq $ 0.99}}
\centerline{\timing{immediately} \responseF{navigateToSafeArea}}

%\begin{itemize}
%    \item requirements for space applications: NASA guide/tech report that Andreas mentioned and do a quick literature search in case we missed anything
%\end{itemize}

For any \fretish{} requirement, both LTL and Probabilistic Computation Tree Logic (PCTL*) semantics are provided by FRET. In this paper, we use the LTL semantics since probabilistic requirements did not appear very often in our dataset but we discuss probabilistic requirements briefly in Section \ref{sec:evaluation}. To the best of our knowledge, although general guidance, standards and space agency processes exist that are related to requirement specification for space missions \cite{Author2015MethodologyComplexSpaceMissions, hirshorn2017nasa,farrell2021evolution,redfield2024verification}, we are not aware of any related work that aims to derive formal specification patterns for autonomous robotic space missions.

\section{Methodology }
\label{sec:methodology}
In this section, we outline the methodology that we followed when assembling a corpus of space requirements to answer the research questions that we define below:\smallskip
\begin{compactitem}
    \item[\textbf{RQ1:}] Are existing specification patterns relevant and sufficient for autonomous space robotic missions?\smallskip
    \item[\textbf{RQ2:}] Can specification patterns for autonomous space robotic missions be expressed using FRET?
\end{compactitem}
\smallskip
\noindent To answer these research questions, we assembled a corpus of 116 requirements for (autonomous) space robotic missions. In total, we reviewed 30 different literature sources to identify and formalise requirements for autonomous space robotics. These sources were found using Google Scholar\footnote{\url{https://scholar.google.com/}} and the NASA Technical Reports Server\footnote{\url{https://ntrs.nasa.gov/}}. Of these sources, 14 were used to derive requirements. A wide range of topics in autonomous space robotics were encountered, including docking, manoeuvring, robotic systems, satellites, and more. These topics were explored by including the search terms ``Requirements OR Verification OR Specification''. Snowballing was used wherever possible. This resulted in the identification of 116 requirements. 

We provide details about where the in-scope papers were published in Table \ref{tab:publications}. There is a diverse set of conference, journal, book and technical report publications in this dataset. There is a mix of venue types ranging from more application-oriented venues to formal methods and requirement engineering conferences. There was some duplication as some papers were extended in more detail as NASA technical reports. However, if the same requirement appeared in multiple sources it was only counted once.

For each requirement that we found in the literature, we recorded: a natural language description, the component/system name and the literature source. We also determined whether they fit into a pre-existing pattern formulation from \cite{Menghi2019SpecPatternsRoboticMissions,Vazquez2024RoboticMissionFRET}. When they did not fit into a pre-existing template, we decided, based on the data, whether a new template should be defined. For all new templates, we provided an LTL and \fretish{} formalisation. For all existing templates, we instantiated the pattern using the \fretish{} templates from \cite{Vazquez2024RoboticMissionFRET}. 

Several examples of the natural language descriptions that we encountered are shown in Table \ref{tab:missionreqs}. Many of these were ambiguous and required additional context from the surveyed papers to truly understand. For example, timing was often omitted in the natural language, so it was mostly inferred based on our understanding of the associated mission described in the relevant paper. The full set of the 116 formalised requirements is in the Appendix and available in our repository\footnote{\url{https://github.com/mariefarrell/spacepatterns}}.

\begin{table}[t]
\centering
\begin{tabular}{|l|l|l|}
\hline
\textbf{Venues}  & \textbf{Type} & \textbf{Ref} \\
\hline
Acta Astronautica & Journal & \cite{Author2015MethodologyComplexSpaceMissions}\\
Autonomy Requirements Engineering for Space Missions & Book & \cite{Vassev2014VerificationAutonomyRequirements}\\
ACM Transactions on Software Engineering and Methodology & Journal & \cite{Crow1996SpaceShuttleFormalizingRequirements} \\
Revista Tecnología en Marcha & Journal & \cite{AuthorCubeSatFlightSoftwareFramework} \\
Space Operations Conference & Conference & \cite{Bualat2018Astrobee} \\
Technical Report from NASA & Technical Report & \cite{Author2023FormalMethodsCaseStudiesDO333,katis2022realizability,agogino2024recommendations, Pressburger2023LiftPlusCruiseFRET}\\
Computer Aided Verification & Conference & \cite{katis2022capture} \\
%Workshop on Formal Methods for Autonomous Systems & Workshop & \cite{Dutle2020ICAROUSOverview}\\ 
Requirements Engineering: Foundation for Software Quality & Conference & \cite{pressburger2023authoring}\\
IEEE Robotics and Automation Magazine  & Journal & \cite{IntBall22024RAIMag}\\
Progress in Aerospace Sciences & Journal & \cite{Pham2014RoboticsOnOrbitServicingReview}\\
%NASA Formal Methods & Conference &\\
Space Science Reviews & Journal & \cite{Mars2020MissionOverview}\\
\hline
\end{tabular}
\caption{Publication venues for the in-scope papers that we found.}% [CHECK THAT ALL OF THE PAPERS CITED IN THE APPENDIX ARE HERE].}
\label{tab:publications}
\end{table}

\begin{table}[t]

\begin{center}
\scalebox{0.9}{
\begin{tabular}{|p{0.8cm}|p{6cm}|p{5cm}|p{2.6cm}|p{1cm}|}
\hline 
\textbf{ID} &
\textbf{Natural Language Description} & \textbf{FRETish} & \textbf{Pattern} & \textbf{Source} \\
\hline \hline
5 & Deploy the parachute using navigated position information once safe parachute deployment velocities have been reached. & \conditionF{if parachutedistance \textless{}= safeparachutedistance} \component{EDL} shall \timing{immediately} \responseF{DeployParachute} &
  \textsc{Triggered \qquad Instant \qquad \qquad Reaction} &
  \cite{Mars2020MissionOverview}  \\ \hline
   9-B&The IM shall be isolated in case of contingency. & \scope{In SAFEMODE} \conditionF{Whenever CONTINGENCY} \component{IM} shall \timing{immediately} \responseF{isolated} &
 \textsc{Modal Instant Reaction}
 & \cite{Author2015MethodologyComplexSpaceMissions} \\
\hline
16 & Autonomously release MMO when the polar orbit is reached.&  \conditionF{Upon polar\_orbit\_reached} \component{BepiColombo} shall \timing{immediately} \responseF{ release\_MMO} & \textsc{Triggered \qquad Instant \qquad\qquad Reaction} &
\cite{Vassev2014VerificationAutonomyRequirements} \\
\hline
   27 &If it is required to know the state of the spacecraft, even during the section of the orbit without a communication link with ground segment, store telemetry data. & \conditionF{If StateRequired \& !CommunitcationLink} \component{CubeSat} shall \timing{immediately} \responseF{StoreData} 
   & \textsc{Triggered \qquad Instant \qquad \qquad Reaction}
   & \cite{AuthorCubeSatFlightSoftwareFramework} \\
\hline
 46 &The SPHERES satellites, however, triangulate their position using infrared/ultrasonic beacons, preventing them from navigating outside the two-meter cube defined by the fixed beacon locations. &  \conditionF{Whenever moving} \component{SPHERES} shall \timing{immediately} \responseF{x\textless{}2 \& y\textless{}2 \& z\textless{}2} 
 & \textsc{Stay-In-Perimeter}
 & \cite{IntBall22024RAIMag}\\
\hline
60-B & A blue ``Aud'' light tells the crew that the microphone is on. &  \scope{In AudioRecording} \component{Astrobee} shall \timing{always} \responseF{BlueAudLED} &
  \textsc{Modal \qquad Maintain Safe Space}
  & \cite{Bualat2018Astrobee} \\
\hline
77-B & Furthermore, when the Int-Ball2 automatically detects that the remaining battery power is low, it returns to the DS for recharging. & \conditionF{Whenever IntBall2Power \textless{}= SafeBattery} \component{IntBall2} shall \timing{at the next timepoint} \responseF{RechargeMode} &
 \textsc{Prompt \qquad \qquad Reaction}
 & \cite{IntBall22024RAIMag} \\
\hline

\end{tabular}}
\end{center}
\caption{Examples of requirements from the space missions that are described in Section \ref{subsec:missionoverview}.}
\label{tab:missionreqs}
\end{table}

\section{Requirement Specification Patterns}
\label{sec:requirements}
In this section, we summarise the space missions that we encountered through our survey, define several new patterns using LTL and FRET, explore the distribution of various patterns (both new and old) throughout our dataset and describe the results of an expert evaluation study that we carried out.

\subsection{Description of Space Missions Surveyed}
\label{subsec:missionoverview}
\begin{figure}[t]
    \centering
    \includegraphics[width=0.8\linewidth]{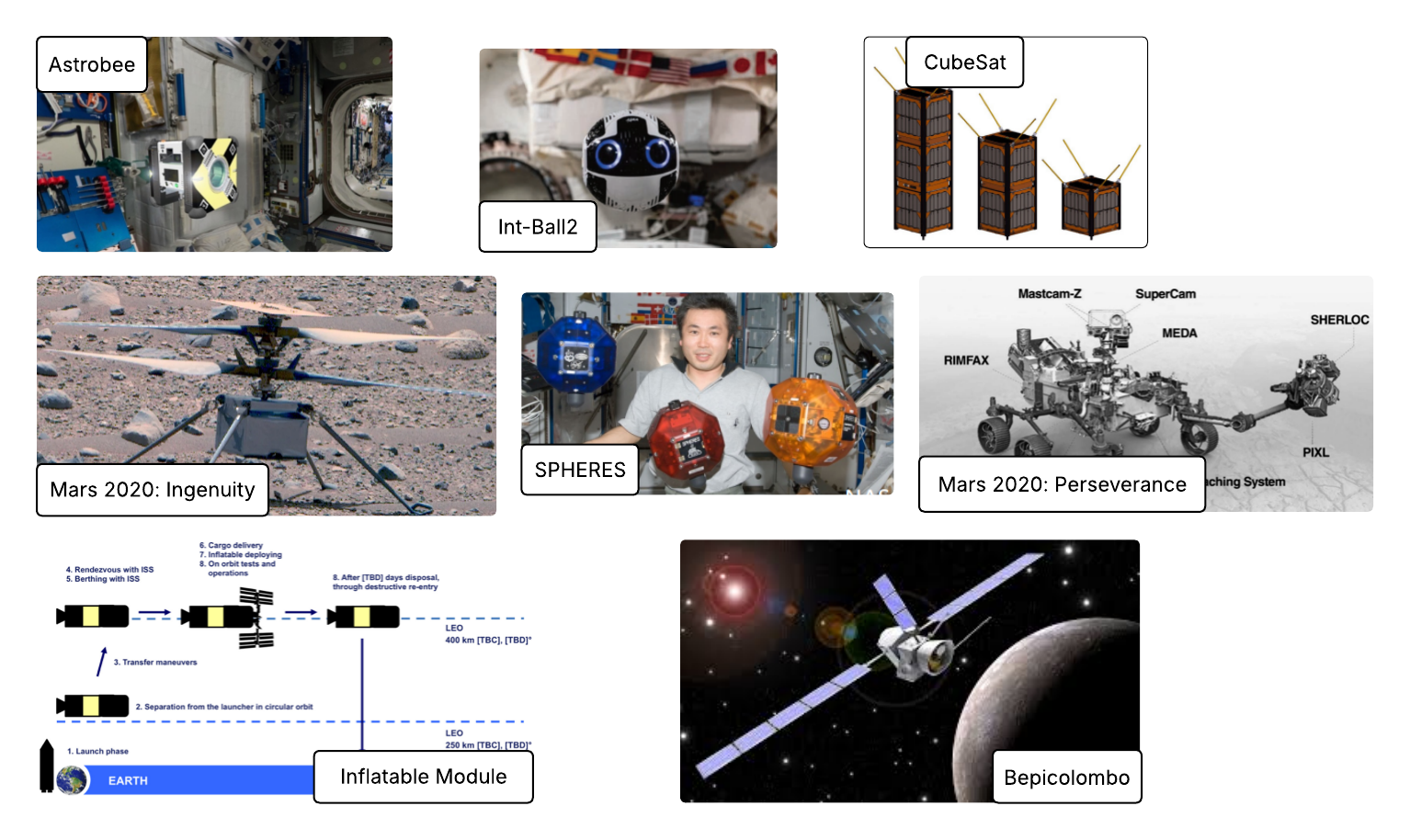}
    \caption{The space missions that are described in Section \ref{subsec:missionoverview}. These images are taken from the surveyed literature, NASA, ESA and JAXA sources.}
    \label{fig:missionimages}
\end{figure}

We describe several of the robotic space missions that were found during our review. Figure \ref{fig:missionimages} illustrates aspects of these systems. We include an example requirement from each of these missions in Table \ref{tab:missionreqs}.

\begin{description}
    \item[\textbf{Astrobee:}] This is a free-flying robot system that is designed to assist astronauts on the International Space Station (ISS). The system comprises free-flying robots, a docking station that provides an Ethernet connection and recharging capabilities and a ground control segment. Astrobee can operate autonomously or under remote control. Astrobee aims to reduce crew workload by performing routine monitoring tasks, handling contingencies and improving productivity onboard the ISS \cite{Bualat2018Astrobee}. Requirement \#60-B in Table \ref{tab:missionreqs}  for Astrobee describes a requirement related to the crew interface for Astrobee.
    
    \item[\textbf{Int-Ball2:}] Like Astrobee, Int-Ball2 is a free-flying robotic system that is designed to assist astronauts on the ISS. It was developed to overcome the shortcomings of the first Int-Ball with enhanced propulsion, markerless navigation and a docking station for autonomous charging. Int-Ball aims to assist astronauts by capturing photos and videos, reducing the need for crew-operated cameras \cite{IntBall22024RAIMag}. Requirement \#77-B in Table \ref{tab:missionreqs} describes a requirement related to when Int-Ball2 should autonomously recharge its battery.
    
    \item[\textbf{Mars 2020:}] This mission saw the deployment of a rover named `\textit{Perseverance}' and a helicopter called `\textit{Ingenuity}' to a chosen Martian crater called Jezero. The mission had 2 goals: (1) finding evidence of microbial life and (2) collecting samples for analysis on Earth. The design of Perseverance builds on the successful design of the \textit{Curiosity} rover with new science instruments for improved analysis and collection. Perseverance also tests new technologies to support future exploration of Mars \cite{Mars2020MissionOverview}. Requirement \#5 in Table \ref{tab:missionreqs} is specific to the Entry, Descent and Landing (EDL) subsystem of the Perseverance rover and specifies when the parachute should be deployed.
    
    \item[\textbf{Inflatable Module:}] This is a case study that presents an in-orbit validation mission of an Inflatable Module (IM). The mission involves launching a spacecraft that docks with the ISS, delivering supplies and testing the IM before disposal through a controlled destructive re-entry. The system operates in five modes across two variants and ten mission phases \cite{Author2015MethodologyComplexSpaceMissions}. Requirement \#9-B in Table \ref{tab:missionreqs} specifies how the IM should be isolated while operating under specific conditions in safe mode.
    
    \item[\textbf{BepiColombo:}] This is a mission to study and research Mercury. The mission is carried out by two satellites: Mercury Planetary Orbiter (MPO) and Mercury Magnetospheric Orbiter (MMO). These two satellites are joined together by the BepiColombo transfer module. The BepiColombo module will provide propulsion, power, and communications while travelling to Mercury. Upon reaching Mercury, the transfer module will release the MPO and the MMO, allowing their independent observation of Mercury's surface, exosphere and composition \cite{Vassev2014VerificationAutonomyRequirements}. Requirement \#16 in Table \ref{tab:missionreqs} describes  a requirement on the MMO release operation of BepiColombo.

    \item[\textbf{CubeSat:}] CubeSat satellites are a class of miniaturised satellites with the dimensions 10cm x 10cm x 10cm or 1U. These small satellites can be configured with one another to create larger CubeSats; for example, a 2U CubeSat would be 10cm x 10cm x 20cm, and a 3U CubeSat would be 10cm x 10cm x 30cm. They provide an accessible and low-cost platform for research and monitoring in space, making them widely used by universities, startups, and space agencies \cite{AuthorCubeSatFlightSoftwareFramework}. In Table \ref{tab:missionreqs}, requirement \#27 describes a communication specific behaviour of the CubeSat mission.

    \item[\textbf{SPHERES:}] SPHERES (Synchronized Position Hold Engage and Reorient Experimental Satellites) is a free-flying system deployed in the ISS. Mainly used as a zero-g research platform, SPHERES has been used in the validation of metrology, formation flight, and autonomy algorithms. However, its design is outdated, relying on CO$_2$ for propulsion and requiring significant support from the ISS crew to operate \cite{SaenzOtero2008SPHERES}\cite{Bualat2018Astrobee}. Requirement \#46 in Table \ref{tab:missionreqs} captures a property related to localisation (that we will discuss in Section \ref{subsec:newpatterns}).

\end{description}

This section provided an overview of the main types of space systems that were explored in our literature review. Included in the literature review were some more generic, relevant aerospace systems; they were omitted here for brevity but they are included in the full list in the Appendix and our repository.

\subsection{New Patterns with LTL and FRET Formalisation}
\label{subsec:newpatterns}
\begin{figure}[t]
    \centering
    \includegraphics[width=\linewidth]{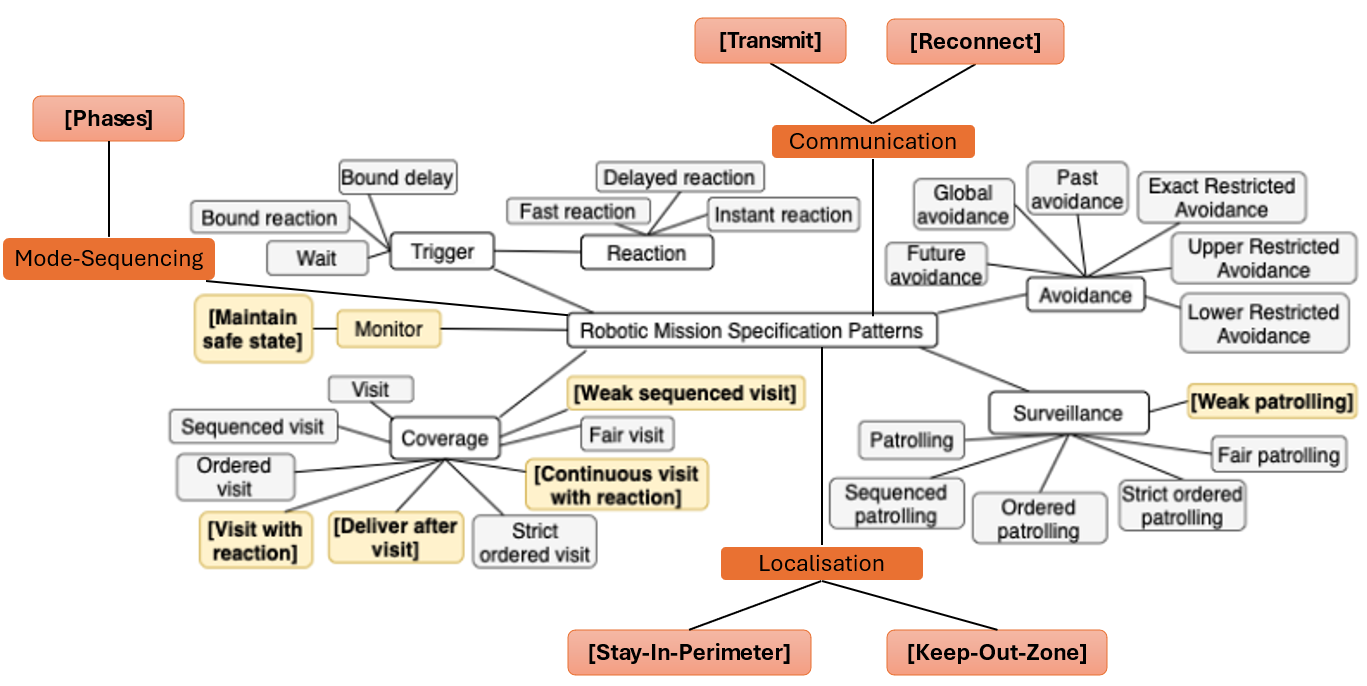}
    \caption{Robotic mission patterns, those with a white/grey background were proposed in \cite{Menghi2019SpecPatternsRoboticMissions}, those with a yellow background were proposed in \cite{Vazquez2024RoboticMissionFRET} and our new patterns are shown with an orange background.}
    \label{fig:spacepatterns}
\end{figure}

Based on our analysis of the literature, we identified 5 new patterns (\textsc{Phases}, \textsc{Transmit}, \textsc{Reconnect}, \textsc{Stay-In-Perimeter}, \textsc{Keep-Out-Zone}) that fall into three distinct categories: \textbf{Mode Sequencing}, \textbf{Communication} and \textbf{Localisation}. These are illustrated in orange in Figure \ref{fig:spacepatterns}. We also found several variations of the existing patterns from \cite{Menghi2019SpecPatternsRoboticMissions,Vazquez2024RoboticMissionFRET} that we discuss in Section \ref{subsec:metrics}. \medskip

\noindent\textbf{Mode Sequencing} is an important aspect of many cyber-physical systems and this was also apparent in the space applications that we explored. In this category we identified and formalised one pattern as discussed below. \smallskip

\noindent\textsc{Pattern 1 (Phases):} We found several space applications such as Mars 2020 \cite{Mars2020MissionOverview} and BepiColumbo \cite{Vassev2014VerificationAutonomyRequirements}, that incorporate phases during operating modes. Intuitively, the phase evolution requirements that we observed in the literature review described the linear transition from one phase to another as individual phases complete. In the work that we considered, there was normally a terminal phase in these requirements. We describe this using LTL as follows.  Let $P = {p_1,\ldots, p_n}$ be a set of phases and $C = {c_1, \ldots, c_n}$ be a set of conditions that are caused by phases and cause phases to begin. Each $c_i$ is produced by phase $p_i$ and will trigger the start of phase $p_{i+1}$. 

\centerline{$((G (((! p_1) \land (X p_1)) \rightarrow (X (F c_1)))) \land (p_1 \rightarrow (F c_1))) $}
\centerline{$\land \ ((G (((! c_1) \land (X c_1)) \rightarrow (X (X p_2)))) \land (c_1 \rightarrow (X p_2))) \ldots
$}

We use multiple\footnote{Following \cite{Vazquez2024RoboticMissionFRET}, we use `\texttt{+}' to indicate the logical conjunction of multiple requirements.} \fretish{} requirements to describe this kind of phase evolution as follows:

\centerline{\conditionF{upon p1} \component{System} shall \timing{eventually} \responseF{c1}}

\centerline{\texttt{+} \conditionF{upon c1} \component{System} shall \timing{at the next timepoint} \responseF{p2}}

\centerline{\texttt{+} \conditionF{upon p2} \component{System} shall \timing{eventually} \responseF{c2}}

We can instantiate this pattern with part of a requirement from the Inflatable case study (also illustrated in Figure \ref{fig:missionimages}) \cite{Author2015MethodologyComplexSpaceMissions}:
\begin{quote}
\textit{`The launch phase begins with lift-off and ends at burn out. The Separation phase begins with burn out, leading to transfer orbit insertion. During transfer, the spacecraft moves toward the Cygnus arrival near the ISS. Finally, the rendezvous phase covers the approach and capture by the robotic arm.''}
\end{quote}
\centerline{\conditionF{upon LaunchPhase} \component{System} shall \timing{eventually} \responseF{burnout}}

\centerline{\texttt{+} \conditionF{upon burnout} \component{System} shall \timing{at the next timepoint} \responseF{SeperationPhase}}

\centerline{\texttt{+} \conditionF{upon SeperationPhase} \component{System} shall \timing{eventually} \responseF{transferorbit}}

This pattern captures the concept of phases that is ubiquitous in many space systems. It ensures that higher level requirements such as phases or steps flow correctly one after the other. This pattern is quite novel when compared to the pre-existing catalogue of 28 requirements \cite{Menghi2019SpecPatternsRoboticMissions,Vazquez2024RoboticMissionFRET}. Structurally, it shares some similarities with the LTL from the \textsc{Sequenced Visit} patterns in \cite{Menghi2019SpecPatternsRoboticMissions}. However, the intent of phase evolution is significantly different from visiting specific locations. \medskip

\noindent\textbf{Communication} is an essential element of space systems since they operate at vast distances away from physical human contact. We found the following two patterns that fit into this category. \smallskip

\noindent\textsc{Pattern 2 (Transmit):} Data can be transmitted when a systems' connection requirements are met. Let $c_i$ represent the necessary connections, $d_i$ represents the groups of data to be transmitted and $T$ denotes the transmitting protocol. If all of the necessary connections are present and there is data that needs to be sent, the transmission will continue until all of the data is sent. We express this in LTL as follows
\centerline{$G\big( ((\bigwedge_{i=1}^{n} c_i) \land (\bigwedge_{i=1}^{n}\neg d_i)) \ \rightarrow\ (T \ \mathcal{W} \bigwedge_{i=1}^{n}d_i)) \big)$}
Many space systems transmit collections of data to ground control or other systems. Transmission is necessary for communication and receiving/sending mission critical information. Some systems rely on the connection of more than one device hence why this pattern covers multiple connections. This pattern incorporates part of the \textsc{Wait} pattern from the 2019 catalogue as the structure of the consequent \cite{Menghi2019SpecPatternsRoboticMissions}.

We express this pattern in \fretish{} as follows. 

\centerline{\conditionF{Whenever c \& !d} \component{System} shall \timing{until d} \responseF{T}}

We can instantiate this pattern with a requirement from Astrobee \cite{Bualat2018Astrobee}:
\begin{quote}
\textit{``After a sortie, Astrobee transfers large files through a hard-wired Ethernet connection with its dock.''}
\end{quote}
\centerline{\conditionF{Whenever Ethernet \& ISSConnection \& !LargeFiles} \component{Astrobee} shall} 
\centerline{\timing{until LargeFiles} \responseF{transmit}}
Note that the ISSConnection condition was implicit in the requirement, it was derived from the documentation. \medskip

\noindent\textsc{Pattern 3 (Reconnect):} Connection can be lost and it is essential to restore that connection. Here, $k_i$ represents connections and $R$ represents the reconnection protocol. If one necessary connection is down then reconnection will occur until all necessary connections are formed.

\centerline{$G((\bigvee_{i=1}^{n}\neg k_i) \rightarrow (\text{R}\ \mathcal{W} \bigwedge_{i=1}^{n}k_i))$}

Due to the nature of connections in space, it is common for loss of signal to occur. Many of those systems that deal with transmitting data are bound to need to reconnect. As a result, this pattern is widely applicable to many systems. Examples of connections: ISS, Ethernet, satellites and ground control. Similar to the \textsc{Transmit} pattern, this one also uses part of the \textsc{Wait} pattern from the 2019 catalogue as the structure of the consequent \cite{Menghi2019SpecPatternsRoboticMissions}. We express this using FRET as follows.

\centerline{\conditionF{Whenever !k1 | !k2} \component{System} shall \timing{until (k1 \& k2)} \responseF{R}}

We can instantiate this pattern with a requirement from Astobee \cite{Bualat2018Astrobee}:

\begin{quote}
\textit{``The space-to-ground network is subject to frequent losses of signal. After loss-of-signal (LOS) Astrobee shall attempt to reconnect.''}
\end{quote}
\centerline{\conditionF{Whenever !GroundSignal | !ISSConnection} \component{Astrobee} shall}
\centerline{\timing{until (ISSConnection \& GroundSignal)} \responseF{Reconnect}} 

We note that both \textbf{Communication} patterns use weak, rather than strong, until. This makes implicit the ability for the communication to fail, as is common in space applications. Specifically, if we had used strong until then, we would require that the transmission (resp. reconnection) protocols eventually succeed, which may not be possible in practice. \medskip

\noindent\textbf{Localisation} is difficult for space applications since they can't simply rely on GPS and the environment within which they operate doesn't always contain easily recognisable features such as landmarks. That said, operating within safe boundaries was a concern in the papers that we reviewed. As a result, we identified the following two patterns.\smallskip

\noindent\textsc{Pattern 4 (Stay-In-Perimeter):} This pattern is used in systems that are required to stay within certain boundaries when specific actions are executed. In the LTL formula below, $a$ represents actions such as movement, approach and velocity matching. Then,  $l_i$ represents the areas/boundaries that the system must remain within.

\centerline{$G(a \rightarrow (\bigwedge_{i=1}^{n} l_i))$}

During the movement/actions of some autonomous space robotics, robots are restricted to only moving/acting inside the perimeter. These safety restrictions are necessary for dynamic or uncontrolled environments for many moving systems. This pattern is structurally similar to the \textsc{Instant Reaction} pattern with a similar intent to the \textsc{Maintain Safe Space} pattern from the previous catalogue \cite{Vazquez2024RoboticMissionFRET}. We express this using FRET as follows.

\centerline{\conditionF{Whenever a} \component{System} shall \timing{immediately} \responseF{l1 \& l2}}

We can instantiate this pattern with a requirement from Int-Ball \cite{IntBall22024RAIMag}:
\begin{quote}
    \textit{``Int-Ball cannot operate without a direct line-of-sight (LOS) to its markers.''}
\end{quote}

\centerline{\conditionF{Whenever operating} \component{IntBall2} shall \timing{immediately} \responseF{LOS1 \& LOS2}} \medskip

\noindent\textsc{Pattern 5 (Keep-Out-Zone):} This is essentially the dual of the previous pattern. We found several cases where systems should avoid certain areas while executing certain actions. As before, $a$ represents actions and $l_i$ represents the areas that the system must avoid. 

\centerline{$G(a \rightarrow (\bigwedge_{i=1}^{n}\neg l_i))$}

Unlike the previous pattern, this one focuses on avoiding areas, rather than remaining in certain areas. This is essential for avoiding dangerous areas or critically important areas.  We formalise this pattern in FRET as follows. 

\centerline{\conditionF{Whenever a} \component{System} shall \timing{immediately} \responseF{!l1 \& !l2}}

We can instantiate this pattern with a requirement from Astrobee \cite{Bualat2018Astrobee}:
\begin{quote}
    \textit{``Astrobee’s navigation and control systems understand the concept of a keep-out zone (KOZ). KOZs are defined as areas where Astrobee is not allowed to fly.''}
\end{quote}

\centerline{\conditionF{Whenever moving} \component{Astrobee} shall \timing{immediately} \responseF{!KOZ1 \& !KOZ2}} \medskip

This pattern is very similar to the Future Avoidance pattern from \cite{Menghi2019SpecPatternsRoboticMissions} which is expressed in \fretish{}  \cite{Vazquez2024RoboticMissionFRET} as, for example ``\conditionF{whenever a} \component{Robot} shall \timing{never} \responseF{l1}''. This gives the LTL formula $G (a \rightarrow (G (\lnot l_1)))$. Our version is more strict than this and the LTL formula that we use actually encompasses the original \textsc{Future Avoidance} pattern from \cite{Menghi2019SpecPatternsRoboticMissions}. In their pattern, the robot is not prohibited from entering $l_1$ at the same time instant that $a$ is true. Our pattern prohibits this behaviour which is what, we believe, was the original intention of  the requirements that we saw in this category. 

%ADD SOMETHING ABOUT THESE BEING MORE WIDELY APPLICABLE, NOT JUST FOR SPACE MISSIONS.

%Although these patterns were derived from space requirements,

\begin{table}[!ht]
    \centering
    \scalebox{0.9}{
    \begin{tabular}{|p{0.8cm}|p{6cm}|p{5.35cm}|p{2.4cm}|p{1cm}|}
    \hline
        \textbf{ID} &\textbf{Natural Language Description} & \textbf{\fretish{} }& \textbf{Pattern}  & \textbf{Source} \\\hline\hline
        2 &  PIXL’s hexapod can compensate for X-Y drift if it is found to exceed a pre-defined threshold. &
  \scope{In Experiment} \conditionF{whenever !inthreshold} \component{Rover} shall \timing{eventually} \responseF{inthreshold} & \textsc{Modal} \qquad \textsc{Delayed Reaction} &
  \cite{Mars2020MissionOverview} \\ \hline
  3 &   In order to ensure that PIXL’s XRF and OFS subsystems are behaving in the expected manner, the instrument’s performance is periodically checked by measuring the onboard calibration target, and then comparing the results of those measurements against pre-flight measurements of those standards. & \conditionF{whenever timetocheck=1} \component{PIXL} shall \timing{at the next timepoint} \responseF{check} &
  \textsc{Prompt \qquad Reaction} &
  \cite{Mars2020MissionOverview} \\ \hline
  7& Only once the rover is safely on the martian surface will flight software command the preparation and downlink of EDL Camera images and microphone data. &
  \conditionF{Upon SafeLanding} \component{EDL} shall \timing{at the next timepoint} \responseF{Preparation \& DownLink} &
  \textsc{Triggered Prompt \qquad Reaction} &
\cite{Mars2020MissionOverview} \\ \hline
10-A &
  Check mode\: all components necessary to check system's health before starting the tests are active. &
  \scope{In CHECKMODE} \component{IMCOMPONENT} shall \timing{before StartingTests} \responseF{necessary\_check\_components\_active} &
  \textsc{Modal Scheduling} &
   \cite{Author2015MethodologyComplexSpaceMissions}  \\ \hline
         11-B &
    Data are transmitted to SM to be elaborated, then transmitted to ISS and eventually to Ground Segment. &
  \scope{In NOMINALTESTINGMODE} \conditionF{whenever SMConnection \&  ISSConnection \& GroundSegment \& !files} \component{IM} shall \timing{until files} \responseF{transmit} &
  \textsc{Modal Transmit} &
   \cite{Author2015MethodologyComplexSpaceMissions}  \\ \hline
   29 &At least one side shall be the pilot flying side. &
  \component{FGS} shall \timing{always} \responseF{PilotFlying \textless{}= 1} &
  \textsc{Semi-Autonomous} &
  \cite{Author2023FormalMethodsCaseStudiesDO333} \\ \hline
   36 &  The probability that the aircraft leaves the taxiway, i.e., |cte| \textgreater 8 meters, shall be extremely low. &
  \component{Aircraft} shall \probability{with probability \textless{}= 0.001} \timing{eventually} \responseF{absReal(cte) \textgreater 8} &
  \textsc{Probabilistic Maintain Safe Space} &
  \cite{agogino2024recommendations}\\ \hline
  41 & The vehicle remains in the thrust-borne mode (TB) as long as kgs \textless{}= 20.0 knots and Hover Control (HC) mode is selected. &
  \scope{In HCmode} \conditionF{whenever TBMode \& Kgs \textless{}= 20} \component{LPC} shall \timing{always} \responseF{TBmode} &
  \textsc{Maintain Mode In \qquad Hierarchy} &
  \cite{Pressburger2023LiftPlusCruiseFRET} \\ \hline
  56 &  If multiple Astrobees are active, the Control Station displays the positions of all of the Astrobees so that the operators are aware of the other activities and can avoid collisions. &
  \conditionF{Whenever numberofAtrobees \textgreater{} 1} \component{CS} shall \timing{immediately} \responseF{DisplayALL} &
  \textsc{Instant \qquad Reaction} &
  \cite{Bualat2018Astrobee} \\ \hline
  77-A &
 However, if the vSLAM output remains unavailable for an extended period, the robot rotates in place until the feature points detected in the current view align with those in the stored map. &
  \conditionF{whenever vSLAMUnavailable} \component{IntBall2} shall \timing{until FeaturePointDetected} \responseF{RotateProtocol} + \conditionF{whenever vSLAMUnavailable} \component{IntBall2} shall \timing{eventually} \responseF{FeaturePointDetected} &
  \textsc{Conditional Wait} &
 \cite{IntBall22024RAIMag} \\ \hline
    \end{tabular}}
\caption{Examples of requirements that use existing patterns, pattern variants and potential new patterns.}
    \label{tab:reqdiscussion}
\end{table}

\subsection{Requirement Pattern Metrics and Analysis}
\label{subsec:metrics}
We include a representative set of requirements in Table \ref{tab:reqdiscussion} for discussion purposes. The full list of requirements is in the Appendix for the interested reader. We summarise the distribution of the requirement pattern categories that were used in Table \ref{tab:metrics}. We use a white background for patterns that were present in the pre-existing sets from \cite{Menghi2019SpecPatternsRoboticMissions,Vazquez2024RoboticMissionFRET}, an orange background for our newly defined patterns, a blue background for variants of the pre-existing and new patterns, and a green background for potential new patterns that require further exploration.

Of our 116 requirements, 41 used pre-existing patterns. The majority, 34, of these used different kinds of reaction patterns: \textsc{Instant Reaction, Prompt Reaction, Delayed Reaction} and \textsc{Visit with Reaction}. The other existing patterns that were used were the \textsc{Maintain Safe Space} and \textsc{Wait} patterns. Examples of these instantiated requirement patterns can be seen in requirements \#3 (\textsc{Prompt Reaction}) and \#56 (\textsc{Instant Reaction}) in Table \ref{tab:reqdiscussion}.

Most of our requirements (52 out of 116) used variations of  pre-existing patterns. These variations were not seen in \cite{Menghi2019SpecPatternsRoboticMissions,Vazquez2024RoboticMissionFRET}. We also had 1 instance of a variation of our new \textsc{Transmit} pattern. In total, we identified 12 distinct pattern variations and most of these (7 out of 12) were variations that involved system modes through the scope field in \fretish{}. For example, we see modal versions of several patterns in Table \ref{tab:reqdiscussion}, including \#2 (\textsc{Modal Delayed Reaction}) and \#11-B (\textsc{Modal Transmit}). 

Other pattern variations in Table \ref{tab:reqdiscussion} include what we label as \textsc{Triggered} versions of existing patterns. For example, requirement \#7 in Table \ref{tab:reqdiscussion} shows an instance of the \textsc{Triggered Prompt Reaction} pattern. The idea of trigger conditions is part of the formalisation in \fretish{} where conditions can either be triggers (using \conditionF{if}, \conditionF{upon}, etc.) or holding (using \conditionF{whenever}). We saw a mix of these condition types throughout our formalisation process and FRET forced us to consider which condition should be used for each requirement. The difference is that for a trigger, the response must hold \conditionF{upon} a condition becoming true. A holding condition means that the response holds \conditionF{whenever} the condition is true. Here, the use of FRET helped us to consider subtle aspects of the requirements.

We identified 3 potential new patterns that require further investigation. The first of these, \textsc{Maintain Mode In Hierarchy}, was found in one of the more generic aerospace requirements for the Lift Plus Cruise study \cite{Pressburger2023LiftPlusCruiseFRET}. Essentially, the system must maintain a mode within another system mode (requirement \#41 in Table \ref{tab:reqdiscussion}). We didn't see this in other system requirements however, we did have phase updates within system modes as part of the IM mission \cite{Author2015MethodologyComplexSpaceMissions}. We encoded these using the \textsc{Phases} pattern but we would need to extend this further to capture the behaviour of \textsc{Maintain Mode In Hierarchy}. Since we only found one instance of it and the natural language was ambiguous, we decided to leave this as a potential pattern that might be returned to in an expanded literature review to determine whether it constitutes a pattern in its own right.

The potential pattern called \textsc{Modal Scheduling} was also found in the IM system requirements \cite{Author2015MethodologyComplexSpaceMissions}. Requirement \#10-A in Table \ref{tab:reqdiscussion} specifies that various checks should be carried out \textit{before} the system tests are executed. We had not seen this pattern elsewhere but we do think that with an expanded literature review that it would present itself as a pattern in the future. Similarly, we found 2 instances of a \textsc{Semi-Autonomous} pattern in the more general aerospace use case in \cite{Author2023FormalMethodsCaseStudiesDO333}. These requirements specify which side of the aircraft is the pilot flying side, which is autonomously controlled and how to switch between them, for example requirement \#29 in Table \ref{tab:reqdiscussion}. We believe that this kind of requirement will become more prevalent as more semi-autonomous systems are deployed. However, we found the natural language to be quite vague, so we have left this one for further exploration.

It was clear that system modes played an important role in space systems that was not as apparent in previous work \cite{Menghi2019SpecPatternsRoboticMissions,Vazquez2024RoboticMissionFRET}. In our new patterns, most instances (11 out of 18) corresponded to our newly defined \textsc{Phases} pattern which is used to specify system phase or mode evolution. Our \textbf{Communication} (respectively \textbf{Localisation}) patterns accounted for 4 out of 18 (respectively 3 out of 18) instances.

\begin{table}[t]
\centering

\scalebox{0.86}{\begin{tabular}{|p{6.67cm}|c||p{6.67cm}|c|}
\hline
\textbf{Pattern}               & \textbf{Frequency} & \textbf{Pattern}            & \textbf{Frequency}      \\ \hline\hline
\cellcolor{blue!10} \textsc{Conditional Wait}                  & \cellcolor{blue!10}1      &             
\textsc{Delayed Reaction}                  & 9                       \\ \hline
\textsc{Instant Reaction}                  & 14                      &
\cellcolor{orange!20}\textsc{Keep-Out-Zone}                     & \cellcolor{orange!20}1                       \\ \hline
\cellcolor{green!30} \textsc{Maintain Mode In Hierarchy}        & \cellcolor{green!30}1                       &
\textsc{Maintain Safe Space}               & 5                       \\ \hline                 
\cellcolor{blue!10}\textsc{Modal Delayed Reaction}            &\cellcolor{blue!10}15                      &
\cellcolor{blue!10}\textsc{Modal Instant Reaction}            & \cellcolor{blue!10}1                       \\ \hline
\cellcolor{blue!10}\textsc{Modal Maintain Safe Space}         & \cellcolor{blue!10}13                      &
\cellcolor{blue!10}\textsc{Modal Prompt Reaction}             & \cellcolor{blue!10}1                       \\ \hline
\cellcolor{blue!10}\textsc{Modal Reaction}                    & \cellcolor{blue!10}2                       &
\cellcolor{green!30}\textsc{Modal Scheduling}                  & \cellcolor{green!30}1                       \\ \hline
\cellcolor{blue!10}\textsc{Modal Transmit}                    & \cellcolor{blue!10}1                      &
\cellcolor{blue!10}\textsc{Modal Triggered Instant Reaction}  & \cellcolor{blue!10}2                       \\ \hline
\cellcolor{orange!20}\textsc{Phases}                            & \cellcolor{orange!20}11                      &
\cellcolor{blue!10}\textsc{Probabilistic Maintain Safe Space} & \cellcolor{blue!10}2                       \\ \hline
\textsc{Prompt Reaction}                   & 9                       &
\cellcolor{orange!20}\textsc{Reconnect}                         & \cellcolor{orange!20}1                       \\ \hline
\cellcolor{green!30} \textsc{Semi-Autonomous}                   & \cellcolor{green!30}2                       &
\cellcolor{orange!20}\textsc{Stay-In-Perimeter}                 & \cellcolor{orange!20}2                       \\ \hline
\cellcolor{orange!20}\textsc{Transmit}                          & \cellcolor{orange!20}3                       &
\cellcolor{blue!10}\textsc{Triggered Delayed Reaction}        & \cellcolor{blue!10}1                       \\ \hline
\cellcolor{blue!10}\textsc{Triggered Instant Reaction}        & \cellcolor{blue!10}12                      &
\cellcolor{blue!10}\textsc{Triggered Prompt Reaction}         & \cellcolor{blue!10}2                       \\ \hline
\textsc{Visit with Reaction}               & 2                       &
\textsc{Wait}                              & 2                       \\ \hline

%Grand Total                       & 116                 \\ \hline
\end{tabular}}
\caption{Distribution of patterns throughout the requirements that we found. We use a white background for patterns that were present in the pre-existing sets from \cite{Menghi2019SpecPatternsRoboticMissions,Vazquez2024RoboticMissionFRET}, an orange background for our newly defined patterns, a blue background for variants of the pre-existing and new patterns and a green background for potential new patterns that require further exploration.}
\label{tab:metrics}

\end{table}

\subsection{Expert Evaluation}
\label{sec:evaluation}
We conducted an initial evaluation of our newly identified patterns with experts who have experience working on space robotics as well as similar systems in other domains, including nuclear and aerospace. This comprised 1 academic from the University of Manchester, 1 researcher from an industrial robotics engineering company and 3 experts from governmentally funded space organisations (NASA, JAXA and Satellite Applications Catapult). 

We interviewed each of these participants using a set of slides to guide the discussion\footnote{The slides are available in our repository: \url{https://github.com/mariefarrell/spacepatterns}}. The slides contained the new patterns, an example of each in LTL and \fretish{} and a series of questions for the participants. These questions asked whether they had seen this pattern in use before, were there any variants that we should consider, could they provide us with example uses from their experience and more general feedback. We interviewed most of the participants individually, grouping two of them together to facilitate busy schedules. The pair that were grouped shared a similar experience working on the same project in recent years. One participant provided answers and feedback via email because it was not possible to schedule an interview at the time. They were given the same set of slides as the other participants. We summarise their feedback on a per-pattern basis below and provide overall reflections.

\begin{description}
    \item[\textsc{Pattern 1 (Phases):}] In general, the experts had encountered requirements that specified phase or mode transitions, though not usually hierarchical modes and failures, which is what we found in \cite{Pressburger2023LiftPlusCruiseFRET}. This pattern had been previously encountered by the experts in aerospace and nuclear scenarios, including unmanned aircraft traffic management\footnote{\url{https://store.astm.org/f3548-21.html}}, (nuclear) operational sequences, Earth observation satellites, satellite manoeuvring and planetary rovers (entry, descent, landing, calibration and connection establishment). One expert remarked that there could be an intermediate ``changing phases'' step since phase transitions are not always instantaneous in practice. Further, complex systems may have branching phases to enable failure modes, which is not something that we had observed in our dataset but this point was echoed by several experts.
    \item[\textsc{Pattern 2 (Transmit):}] The experts had encountered this pattern in communications-related requirements but there are further variants that are worth consideration. Specifically, this requirement does not capture a communication being instantiated, that the data is ready to send or account for the different types of communication that may be used such as Ethernet, radio, etc. As such, there are likely more low-level details that need to be instantiated  to use this requirement pattern in practical systems. In a related astronaut-rover use case \cite{webster2020formal}, this pattern would have been split into (1) maintain connection with the astronaut and (2) transmit data. Some experts noticed this requirement in existing use cases related to Earth observation satellites. In these cases, there would likely be more fine-grained transmission timings that might also be included. For example, transmissions would be limited to take a specified number of seconds.
    
    \item[\textsc{Pattern 3 (Reconnect):}] The experts had seen this in systems involving swarms of robots, satellite constellations and a confined space drone. This pattern does not specify how long reconnection might take but, in practice, the system should probably only try and fail for so long before it resorts to another action.  For example, it may need to go into a safe/wait/power-save mode and attempt reconnection again at a later stage. It may even need to autonomously physically reorient or move to a position where it can still harvest solar power to maintain operation, both while it awaits reconnection and as part of its attempt to reconnect. It was noted that this might be split into numerous requirements for each connection type but this would be a domain/application-specific design choice. One expert noted that, in practice, future variants of this pattern may need to capture real-time requirements, which we did not see in our dataset.
    \item[\textsc{Pattern 4 (Stay-In-Perimeter):}] This pattern was frequently encountered by all of our experts in robotics use cases including a fork-lift use case \cite{maoz2016synthesizing}, proximity operations for formation flying satellites, satellites maintaining position, precision targetting, manipulator robots, planetary rovers that must maintain a safe distance from a lander and scenarios that involve robot collectives e.g. swarms, robot soccer, etc. There are also more subtle cases where designers may want to consider a `gravity-well' attractor force to some position. For example, imagine an Astrobee/SPHERES robot that must maintain its (relative) position in space in/to the ISS, so it is commanded to move to that position; if necessary, it might have to move out of the way to avoid an object of space debris, but then return to its position once clear.
    \item[\textsc{Pattern 5 (Keep-Out-Zone):}] Although this is very similar to the previous pattern, it was agreed that the intention was different enough to merit it being a pattern in its own right. This was identified as an especially important pattern for safety systems. This pattern had been observed by experts in transfer vehicles delivering consumables to the ISS, planetary (lunar) rovers that must avoid deep craters, unmanned aircraft traffic management and nuclear scenarios. One expert remarked that it could potentially be adapted to dynamically account for trajectory or even adversary avoidance. In manipulator robotics, there is a possible variant to include a ``slow-down-zone'' for when the robot gets close to a keep-out-zone. This variant was not present in our dataset. Several experts also remarked that, although these dangerous zones should be avoided, it is feasible that a robot may enter such a zone for a short, tolerable period of time before returning to the safe zones. This would require an ability to express real-time requirements, which we do not have in standard LTL or the current version of FRET.  
\end{description}

\noindent\textbf{Overall Comments:} Our experts observed that requirements are rarely completely deterministic in practice, likely incorporating probabilistic uncertainty. Probabilistic requirements can be expressed in  FRET but we only encountered a small number of probabilistic requirements in our dataset (2 out of 116). One of these probabilistic requirements is contained in Table \ref{tab:reqdiscussion} (requirement \#36) which is a probabilistic version of the \textsc{Maintain Safe Space} pattern. Of course, any of the other requirement patterns could be easily made probabilistic, so treat this as a variant of existing rather than a new pattern in Table \ref{tab:metrics}. A comparison of related work on probabilistic specification patterns from \cite{menghi2022mission} comprises an interesting future research direction for this work. Related to uncertainty, several experts noted the lack of a specification or any assumptions about the dynamic and unpredictable environment within which the systems are operating in our patterns. 

A limitation of LTL for these specifications that was observed by multiple experts is the likely need for first-order logical quantifiers to accurately express requirements that are related to multi-robot systems such as swarms, satellite constellations and planetary robot teams. For example, we may want to be able to specify requirements such as all swarm robots must maintain communication between all other swarm robots. Requirement \#56 in Table \ref{tab:reqdiscussion} is of this spirit and would likely be more precise if we were able to use quantifiers in FRET. That said, extending FRET in this way would be a significant (but worthwhile) undertaking as the move to first-order temporal logic would need to be reflected in both the underlying framework and the linked analysis tools.

Almost all of the experts that we interviewed remarked that we did not have patterns that captured error handling, fault tolerance and/or failure recovery. One suggestion was to include patterns related to Failure Mode and Effects Analysis (FMEA) \cite{stamatis2003failure} as a future extension. Related to this would be requirement patterns related to resource management, such as battery consumption. This should be particularly relevant for space applications. We did find one requirement related to the IntBall mission (Requirement \#77-B in the Table \ref{tab:reqdiscussion}) that specifies that the robot should automatically recharge when its battery power is low. However, we did not find other such requirements so we have not designated a separate pattern for this; in fact, it fits the Prompt Reaction pattern from \cite{Menghi2019SpecPatternsRoboticMissions}.

Another interesting point is that we had originally used the \conditionF{whenever} condition in the \textsc{Phases} pattern and this was the version that we presented to the experts. After much discussion, it was agreed that a triggered condition more accurately captured our desired semantics so we changed this to an \conditionF{upon} condition in the pattern shown in Section \ref{sec:requirements}. The main reason for this is that phase
 changes must be triggered (become true from false) at a specific time instant rather than continually holding.

Although we examined these requirements in the context of space missions, several experts remarked that they could be useful in other domains with significant crossover in nuclear applications that they had seen. This is somewhat ironic since, at the start of this work, we set out to find space domain-specific requirements. However, it is useful that we found patterns that can apply across distinct domains.

\section{Discussion: Answering the Research Questions}
\label{sec:discussion}
We now return to the two research questions that this project started with. We begin by discussing our first research question: 

\begin{center}
\noindent\fbox{\parbox{0.98\textwidth}{\textbf{RQ1:} Are existing specification patterns relevant and sufficient for autonomous space robotic \\missions?}}
\end{center}

\vspace{5pt}
\noindent We address this research question through the metrics that we presented in Section \ref{subsec:metrics}. Many of the pre-existing patterns from \cite{Menghi2019SpecPatternsRoboticMissions,Vazquez2024RoboticMissionFRET} were used in our dataset, exhibiting their usefulness for space applications. So for the first part of the question, the answer is yes. The existing patterns are indeed relevant for space applications. However, for the second part of the question, they are not quite sufficient. We needed to define several variants of the existing patterns and 5 new patterns. We also found a small number of requirements that did not clearly fit into any of the patterns that already existed or were defined in this paper so further analysis is required as future work to determine if the potential patterns that we identified become patterns in their own right. This was also reflected in the expert feedback that we received, where the experts identified many variants of the new patterns as important based on their experience in the space sector. These variants included probabilistic, real-time and first-order temporal logic versions of patterns. This area clearly needs more work and an even more detailed study, including pattern use in the development of future space systems, is needed to further expand and more thoroughly evaluate which patterns are most useful for and needed in space applications. In addition, a consideration of how patterns are related would provide an interesting avenue of future work. This would examine whether a hierarchy of patterns would be useful and provide guidance on how patterns could be composed in, for example, a compositional verification framework.

\begin{center}
\noindent\fbox{\textbf{RQ2:} Can specification patterns for autonomous space robotic missions be expressed using FRET?} \end{center} 

\noindent For the second research question, we followed the approach in \cite{Vazquez2024RoboticMissionFRET} and expressed our specification patterns as requirements in FRET. The patterns that we identified were all straightforwardly expressed in \fretish{} using FRET. However, many of the useful variants that were identified by the experts are not currently expressible in \fretish{}. These include first-order temporal formulas as well as those incorporating real-time aspects. These are potential and significant extensions to FRET, which would require further detailed research to enable. Interestingly, the \fretish{} structured natural language influenced our thought process when formalising the requirements, primarily through the use of the \scope{scope}  and \conditionF{condition} fields. Specifically, the \scope{scope} field provided us with a natural way to express many of the requirements that referred to system operating modes. Further, we felt that having to distinguish between trigger and holding conditions forced us to be more precise during formalisation. However, as many of the natural-language requirements were quite vague, it is possible that a domain expert might have opted for the other condition type in some cases. Notably, FRET's use of weak, rather than strong, until was a very positive point as weak until more closely captured the intention of the \textbf{Communication} patterns that we identified. It was useful that this was natural in \fretish{}.

\section{Conclusion and Future Work}
\label{sec:conclude}
As space missions become ever more sophisticated and reliant on autonomous functionality, ensuring correctness through suitable, logically expressed, requirements is essential. In this paper, we sought to examine whether existing robotic specification patterns were applicable in space missions and whether new patterns were necessary for this domain. We answered these questions by providing a literature review which contributed a corpus of 116 formalised requirements from existing space missions. Based on these 116 requirements, we demonstrated that existing patterns were indeed applicable in the space domain. However, we also contributed five new patterns and several pattern variants to adequately capture the requirements that we found. We also identified a number of potential new patterns that require further research. We provide detailed insight into the new patterns through our expert evaluation study which highlighted the utility and limitations of the new patterns that were identified. Our contributions provide a baseline set of space requirements and associated patterns that have spawned several interesting directions of future work including further analysis of potential patterns, real-time and first-order extensions to \fretish{}, and a large-scale validation of these patterns in a space mission. 

Notably, our current baseline requirement patterns for autonomous space robotics could be expanded by exploring space missions in general across multiple space agencies. This broader analysis of space missions would provide new architectural patterns to execute a mission as the patterns used by a given space agency may not be the only way to specify a mission. Moreover, future work could focus on more application-driven user case studies where domain-specific experts would be introduced to the patterns, subsequently apply them using FRET and finally implement the system using e.g. ROS to provide traceability from requirement patterns through to implementations. Such studies would not only validate the patterns in a more practical setting but also encourage adoption by the wider engineering community.

\paragraph{Acknowledgements:}We thank our expert evaluators for their insightful discussion and feedback: Michael Fisher (The University of Manchester), John Brotherhood (Amentum), Tsutomu Kobayashi (JAXA), Andreas Katis (KBR Inc./NASA Ames Research Center) and Dimitris Xydas (Satellite Applications Catapult). This work was supported by the Summer 2025 Google DeepMind Research Ready internship through a partnership with the Royal Academy of Engineering, Google DeepMind and the Hg Foundation. It was also supported by a Royal Academy of Engineering Research Fellowship.

%\clearpage

\bibliographystyle{eptcs}
\bibliography{main}
\clearpage

\appendix
\section{Table of Requirements}
\centering
\begin{longtable}{|c|p{5.5cm}|p{5.5cm}|p{2cm}|p{1cm}|}
\hline
\multicolumn{1}{|r|}{\textbf{ID}} &
  \textbf{Description} &
  \textbf{\fretish} &
  \textbf{Pattern} &
  \textbf{Source} \\ \hline\hline
\multicolumn{1}{|r|}{1} &
  During an experiment: Two context images taken at different times of a PIXL experiment will be compared to detect unplanned movement (drift) of the rover arm likely to arise from temperature changes. &
  \scope{In Experiment} \component{PIXL} shall \timing{eventually} \responseF{Take2Picture \& ComparePicture} &
  \textsc{Modal Reaction} &
  \cite{Mars2020MissionOverview} \\ \hline
\multicolumn{1}{|r|}{2} &
  PIXL’s hexapod can compensate for X-Y drift if it is found to exceed a pre-defined threshold &
  \scope{In Experiment} \conditionF{whenever !inthreshold} \component{Rover} shall \timing{eventually} \responseF{inthreshold} & \textsc{Modal Delayed Reaction} &
  \cite{Mars2020MissionOverview} \\ \hline
\multicolumn{1}{|r|}{3} &
  In order to ensure that PIXL’s XRF and OFS subsystems are behaving in the expected manner, the instrument’s performance is periodically checked by measuring the onboard calibration target, and then comparing the results of those measurements against pre-flight measurements of those standards. & \conditionF{whenever timetocheck=1} \component{PIXL} shall \timing{at the next timepoint} \responseF{check} &
  \textsc{Prompt Reaction} &
  \cite{Mars2020MissionOverview} \\ \hline
\multicolumn{1}{|r|}{4} &
  Once a safe target has been selected, the spacecraft adjusts its trajectory in propulsive powered flight to land at the target. &
  \conditionF{if safetargetlocated} \component{shuttle} shall \timing{immediately} \responseF{AdjustToLand} &
  \textsc{Triggered Instant Reaction} &
  \cite{Mars2020MissionOverview} \\ \hline
\multicolumn{1}{|r|}{5} &
  Deploy the parachute using navigated position information once safe parachute deployment velocities have been reached &
  \conditionF{if parachutedistance \textless{}= safeparachutedistance} \component{EDL} shall \timing{immediately} \responseF{DeployParachute} &
  \textsc{Triggered Instant Reaction} &
  \cite{Mars2020MissionOverview}  \\ \hline
\multicolumn{1}{|r|}{6} &
  LVS begins taking pictures at 4.2 km altitude and matching them up to an onboard map. &
  \conditionF{Whenever Altitude \textless{}= 4.2} \component{LVS} shall \timing{immediately} \responseF{TakePictures \& Match} &
  \textsc{Instant Reaction} &
\cite{Mars2020MissionOverview}  \\ \hline
\multicolumn{1}{|r|}{7} &
  Only once the rover is safely on the martian surface will flight software command the preparation and downlink of EDL Camera images and microphone data. &
  \conditionF{Upon SafeLanding} \component{EDL} shall \timing{at the next timepoint} \responseF{Preparation \& DownLink} &
  \textsc{Triggered Prompt Reaction} &
\cite{Mars2020MissionOverview} \\ \hline
\multicolumn{1}{|r|}{8} &
  Stand-by mode: In IM stand-by mode only components necessary to monitor the system and to survive the external environment shall be active &
  \scope{In STANDBYMODE} \component{IM} shall \timing{always} \responseF{necessary\_components\_only} &
  \textsc{Modal Maintain Safe Space} &
 \cite{Author2015MethodologyComplexSpaceMissions} \\ \hline
9-A &
  Safe mode: In IM safe mode all components are activated at limited level (adopted in case of contingency) &
  \scope{In SAFEMODE} \component{IM} shall \timing{always} \responseF{components\_limited\_level} &
  \textsc{Modal Maintain Safe Space} &
\cite{Author2015MethodologyComplexSpaceMissions}  \\ \hline
9-B &
  The IM shall be isolated in case of contingency. &
  \scope{In SAFEMODE} \conditionF{whenever CONTINGENCY} \component{IM} shall \timing{immediately} \responseF{isolated} &
  \textsc{Modal Instant Reaction} &
\cite{Author2015MethodologyComplexSpaceMissions}  \\ \hline
9-C &
  In case of contingency safe mode is employed &
  \conditionF{Whenever CONTINGENCY} \component{IM} shall \timing{immediately} \responseF{SAFEMODE} &
  \textsc{Instant Reaction} &
\cite{Author2015MethodologyComplexSpaceMissions}  \\ \hline
10-A &
  Check mode\: all components necessary to check system's health before starting the tests are active &
  \scope{In CHECKMODE} \component{IMCOMPONENT} shall \timing{before StartingTests} \responseF{necessary\_check\_components\_active} &
  \textsc{Modal Scheduling} &
   \cite{Author2015MethodologyComplexSpaceMissions}  \\ \hline
10-B &
  If testing is imminent enter check mode &
  \conditionF{whenever TESTING\_IMMINENT} \component{IM} shall \timing{at the next timepoint} \responseF{Checkmode} &
  \textsc{Prompt Reaction} & \cite{Author2015MethodologyComplexSpaceMissions}  \\ \hline
11-A &
  Nominal testing mode: all components necessary to perform tests are active &
  \scope{In NOMINALTESTINGMODE} \component{IMCOMPONENT} shall \timing{always} \responseF{necessary\_testing\_components\_active} &
  \textsc{Modal Maintain Safe Space} &
\cite{Author2015MethodologyComplexSpaceMissions}  \\ \hline
11-B &
  data are transmitted to SM to be elaborated, then transmitted to ISS and eventually to Ground Segment &
  \scope{In NOMINALTESTINGMODE} \conditionF{whenever SMConnection \&  ISSConnection \& GroundSegment \& !files} \component{IM} shall \timing{until files} \responseF{transmit} &
  \textsc{Modal Transmit} &
   \cite{Author2015MethodologyComplexSpaceMissions}  \\ \hline
\multicolumn{1}{|r|}{12} &
  Nominal crew mode: all main functionalities are active and access of the crew to perform visual inspections is allowed. &
  \scope{In NOMINALCREWMODE} \component{IM} shall \timing{always} \responseF{main\_functionalities\_active \& crew\_access} &
  \textsc{Modal Maintain Safe Space} & \cite{Author2015MethodologyComplexSpaceMissions}  \\ \hline
13-A &
  During the Launch Phase the only mode of operation in use is STANDBY mode which should be done when the IM is in a stowed configuration &
  \scope{In LAUNCHPHASE} \conditionF{whenever stowed} \component{IM} shall \timing{eventually} \responseF{STANDBYMODE} &
  \textsc{Modal Delayed Reaction} & \cite{Author2015MethodologyComplexSpaceMissions}  \\ \hline
13-B &
  During the Separation Phase the only mode of operation in use is STANDBY mode which should both be done when the IM is in a stowed configuration &
  \scope{In SEPARATIONPHASE} \conditionF{whenever stowed} \component{IM} shall \timing{eventually} \responseF{STANDBYMODE}  &
  \textsc{Modal Delayed Reaction} &
   \cite{Author2015MethodologyComplexSpaceMissions}  \\ \hline
13-C &
  During the Transfer Phase the only modes of operation in use is STANDBY and SAFE mode which should both be done when the IM is in a stowed configuration &
  \scope{In TRANSFERPHASE} \conditionF{whenever stowed} \component{IM} shall \timing{eventually} \responseF{STANDBYMODE \& SAFE}  &
\textsc{Modal Delayed Reaction} &
\cite{Author2015MethodologyComplexSpaceMissions}  \\ \hline
13-D &
  During the Rendezvous Phase the only modes of operation in use is STANDBY and SAFE mode which should both be done when the IM is in a stowed configuration &
  \scope{In RENDEZVOUSPHASE} \conditionF{whenever stowed} \component{IM} shall \timing{eventually} \responseF{STANDBYMODE \& SAFEMODE}  &
\textsc{Modal Delayed Reaction} &
   \cite{Author2015MethodologyComplexSpaceMissions}  \\ \hline
13-E &
  During the Berthing Phase the only modes of operation in use is STANDBY and SAFE mode which should both be done when the IM is in a stowed configuration &
  \scope{In BERTHINGPHASE} \conditionF{whenever stowed} \component{IM} shall \timing{eventually} \responseF{STANDBYMODE \& SAFEMODE}  &
  \textsc{Modal Delayed Reaction} &
   \cite{Author2015MethodologyComplexSpaceMissions}  \\ \hline
13-F &
  During the Cargo delivery Phase the only modes of operation in use is STANDBY and SAFE mode which should both be done when the IM is in a stowed configuration &
  \scope{In CARGODELIVERYPHASE} \conditionF{whenever stowed} \component{IM} shall \timing{eventually} \responseF{STANDBYMODE \& SAFEMODE} &
   \textsc{Modal Delayed Reaction} & 
   \cite{Author2015MethodologyComplexSpaceMissions}  \\ \hline
13-G &
  During the Inflatable deploying Phase the only modes of operation in use is CHECK, SAFE, NOMINAL TESTING  mode which should be done when the IM is in a stowed or deployed configuration &
  \scope{In INFLATABLEPHASE} \conditionF{whenever (stowed|deployed)} \component{IM} shall \timing{eventually} \responseF{SAFEMODE \& CHECKMODE \& NOMINALTESTINGMODE}  &
\textsc{Modal Delayed Reaction} & 
   \cite{Author2015MethodologyComplexSpaceMissions}  \\ \hline
13-H &
  During the On orbit tests and ops Phase the only modes of operation in use is CHECK, SAFE, NOMINAL TESTING and NOMINAL CREW  mode which should  be done when the IM is in a deployed configuration &
  \scope{In ONORBITSPHASE} \conditionF{whenever deployed} \component{IM} shall \timing{eventually} \responseF{SAFEMODE \& CHECKMODE \& NOMINALCREWMODE \& NOMINALTESTINGMODE}  &
\textsc{Modal Delayed Reaction} &
   \cite{Author2015MethodologyComplexSpaceMissions}  \\ \hline
13-I &
  During the Undocking delivery Phase a mode of operation in use is STANDBY which should both be done when the IM is in a deployed configuration &
  \scope{In UNDOCKINGPHASE} \conditionF{whenever deployed} \component{IM} shall \timing{eventually} \responseF{STANDBY} &
\textsc{Modal Delayed Reaction} &
   \cite{Author2015MethodologyComplexSpaceMissions}  \\ \hline
13-J &
  During the Undocking delivery Phase a mode of operation in use is SAFE which should both be done when the IM is in a deployed or stowed configuration &
\scope{In UNDOCKINGPHASE} \conditionF{whenever (stowed|deployed)} \component{IM} shall \timing{eventually} \responseF{SAFE} &
\textsc{Modal Delayed Reaction} &
   \cite{Author2015MethodologyComplexSpaceMissions}  \\ \hline
13-K &
  During the Destructive re-entry a mode of operation in use is STANDBY which should both be done when the IM is in a deployed configuration &
  \scope{In DESTRUCTIVEPHASE} \conditionF{whenever deployed} \component{IM} shall \timing{eventually} \responseF{STANDBY}  &
\textsc{Modal Delayed Reaction} &
   \cite{Author2015MethodologyComplexSpaceMissions}  \\ \hline
14-A &
  The launch phase begins ends at burn out. &
  \conditionF{Upon LaunchPhase} \component{System} shall \timing{eventually} \responseF{burnout} &
  \textsc{Phases} &
   \cite{Author2015MethodologyComplexSpaceMissions}  \\ \hline
14-B &
  The Separation phase begins with burn out. &
  \conditionF{Upon burnout} \component{System} shall \timing{at the next timepoint} \responseF{SeparationPhase} &
  \textsc{Phases} &
  \cite{Author2015MethodologyComplexSpaceMissions} \\ \hline
14-C &
  The Separation phase ends with transfer orbit insertion.  &
  \conditionF{Upon SeparationPhase} \component{System} shall \timing{eventually} \responseF{orbitinsertion} &
  \textsc{Phases} &
   \cite{Author2015MethodologyComplexSpaceMissions}  \\ \hline
14-D &
  Orbit insertion leads to the beginning of the transfer phase. &
  \conditionF{Upon orbitinsertion} \component{System} shall \timing{at the next timepoint} \responseF{TransferPhase} &
  \textsc{Phases} &
  \cite{Author2015MethodologyComplexSpaceMissions} \\ \hline
14-E &
  During transfer, the spacecraft moves toward the Cygnus arrival near the ISS &
  \conditionF{Upon TransferPhase} \component{System} shall \timing{eventually} \responseF{cyngusarriaval} &
  \textsc{Phases} &
   \cite{Author2015MethodologyComplexSpaceMissions}  \\ \hline
14-F &
  Finally, the rendezvous phase covers the approach. &
  \conditionF{Upon cyngusarriaval} \component{System} shall \timing{at the next timepoint} \responseF{RendezvousPhase} &
  \textsc{Phases} &
  \cite{Author2015MethodologyComplexSpaceMissions} \\ \hline
14-G &
  Finally, the rendezvous phase covers the approach and capture by the robotic arm &
  \conditionF{Upon RendezvousPhase} \component{System} shall \timing{eventually} \responseF{ captureroboticarm} &
  \textsc{Phases} &
   \cite{Author2015MethodologyComplexSpaceMissions}  \\ \hline
\multicolumn{1}{|r|}{15} &
  Autonomously release the SEPM when the right jettison attitude is reached &
  \conditionF{Upon  Currentattitude \textless{}= Rightattitude} \component{ReleaseBepiColombo} shall \timing{immediately} \responseF{ release\_SEPM} &
  \textsc{Triggered Instant Reaction} &
  \cite{Vassev2014VerificationAutonomyRequirements} \\ \hline
\multicolumn{1}{|r|}{16} &
  Autonomously release MMO when the polar orbit is reached &
  \conditionF{Upon polar\_orbit\_reached} \component{BepiColombo} shall \timing{immediately} \responseF{ release\_MMO} &
  \textsc{Triggered Instant Reaction} &
  \cite{Vassev2014VerificationAutonomyRequirements}  \\ \hline
\multicolumn{1}{|r|}{17} &
  Autonomously determine a steering law  &
  \conditionF{whenever operating} \component{BepiColombo} shall \timing{eventually} \responseF{ determine\_steering\_law} &
  \textsc{Prompt Reaction}
  & \cite{Vassev2014VerificationAutonomyRequirements} \\ \hline
\multicolumn{1}{|r|}{18} &
  Use low thrust to achieve capture around Mercury &
  \conditionF{whenever capturing} \component{BepiColombo} shall \timing{immediately} \responseF{low\_thrust} & \textsc{Instant Reaction}
   &
  \cite{Vassev2014VerificationAutonomyRequirements}  \\ \hline
\multicolumn{1}{|r|}{19} &
  Autonomously acquire the escape procedure and use it to leave Mercury if necessary &
  \conditionF{Upon need\_mercury\_escape} \component{BepiColmbo} shall \timing{immediately} \responseF{ acquire\_escape\_procedure \& escape} &
 \textsc{Triggered Instant Reaction} &
  \cite{Vassev2014VerificationAutonomyRequirements}  \\ \hline
20-A &
  Autonomously detect the presence of high solar irradiation &
  \conditionF{whenever high\_solar} \component{BepiColombo} shall \timing{eventually} \responseF{detect} &
  \textsc{Delayed Reaction} &
  \cite{Vassev2014VerificationAutonomyRequirements}  \\ \hline
20-B &
  In case of presence of high solar irradiation the system will be able to shield the electronics by turning them off &
  \conditionF{Whenever solar\_irradiation \textgreater Normal\_solar\_radiation} \component{BepiColmbo} shall \timing{immediately} \responseF{ turn\_off\_electronics} &
  \textsc{Instant Reaction} &
  \cite{Vassev2014VerificationAutonomyRequirements}  \\ \hline
20-C &
  In case of presence of high solar irradiation the system will be able to shield the electronics &
 \conditionF{Whenever solar\_irradiation \textgreater Normal\_solar\_radiation} \component{BepiColmbo} shall \timing{immediately} \responseF{shield\_electronics} &
  \textsc{Instant Reaction} &
  \cite{Vassev2014VerificationAutonomyRequirements}  \\ \hline
20-D &
  Autonomously detect the presence of high solar irradiation and get away if possible, by using chemical propulsion. &
  \conditionF{Whenever solar\_irradiation \textgreater Normal\_solar\_radiation} \component{BepiColmbo} shall \timing{immediately} \responseF{ get\_away\_chemically} &
  \textsc{Instant Reaction} &
  \cite{Vassev2014VerificationAutonomyRequirements}  \\ \hline
\multicolumn{1}{|r|}{21} &
  Autonomously maintain the onboard equipment and the spacecraft structure in proper temperature range. &
  \component{BepiColmbo} shall \timing{always} \responseF{MaintainEquipment \& MaintainTemperature} &
  \textsc{Maintain Safe Space} &
  \cite{Vassev2014VerificationAutonomyRequirements}  \\ \hline
\multicolumn{1}{|r|}{22} &
  The algorithm first selects the vernier jet or the group of primary jets whose acceleration has the largest scalar (dot) product with the desired rotational acceleration vector. &
  \component{SRC} shall \timing{at the next timepoint} \responseF{(SelectFirstJet | SelectPrimaryJets) \& !(SelectFirstJet \& SelectPrimaryJets)} &
  \textsc{Prompt Reaction} &
  \cite{Crow1996SpaceShuttleFormalizingRequirements}\\ \hline
\multicolumn{1}{|r|}{23} &
  If second and third jets are required, they are similarly selected on the basis of the second and third largest scalar products. &
  \conditionF{Whenever SecondJet | ThirdJet} \component{SRC} shall \timing{immediately} \responseF{ SelectNeededJet} &
  \textsc{Instant Reaction} &
  \cite{Crow1996SpaceShuttleFormalizingRequirements} \\ \hline
\multicolumn{1}{|r|}{24-A} &
  If three jets satisfying the given thresholds cannot be found, the algorithm considers pairs, or, as a last resort, single jets &
  \conditionF{Whenever !ThreeJets} \component{SRC} shall \timing{immediately} \responseF{ Considerpairs} &
  \textsc{Instant Reaction} & \cite{Crow1996SpaceShuttleFormalizingRequirements}  \\ \hline
  \multicolumn{1}{|r|}{24-B} &
  If three jets satisfying the given thresholds cannot be found, the algorithm considers pairs, or, as a last resort, single jets &
  \conditionF{Whenever !TwoJets} \component{SRC} shall \timing{immediately} \responseF{ Considersingle} &
  \textsc{Instant Reaction} &
 \cite{Crow1996SpaceShuttleFormalizingRequirements} \\ \hline
\multicolumn{1}{|r|}{25} &
  During the final phase of Shuttle flight, the orbiter must enter a ``heading alignment cylinder'' &
  \scope{In FinalPhase} \component{Orbiter} shall \timing{eventually} \responseF{ HeadingAlignmentsCylinder} &
  \textsc{Modal Delayed Reaction} &
  \cite{Crow1996SpaceShuttleFormalizingRequirements} \\ \hline
\multicolumn{1}{|r|}{26} &
  if three Shuttle main engines fail sequentially or simultaneously begin calculating/commanding safe abort manoeuvres. &
  \conditionF{Upon ThreeEngineFailure} \component{ThreeE\_O} shall \timing{immediately} \responseF{ CalculatePlan \& SafeManoeuvres} &
  \textsc{Triggered Instant Reaction} &
  \cite{Crow1996SpaceShuttleFormalizingRequirements} \\ \hline
\multicolumn{1}{|r|}{27} &
   If it is required to know the state of the spacecraft, even during the section of the orbit without a communication link with ground segment, store telemetry data.  &
  \conditionF{If StateRequired \& !CommunitcationLink} \component{CubeSat} shall \timing{immediately} \responseF{StoreData} &
  \textsc{Triggered Instant Reaction} &
  \cite{AuthorCubeSatFlightSoftwareFramework} \\ \hline
28-A &
  in charge of providing the ground segment with telemetry data about the state and health of the spacecraft, therefore this service shall be able to automatically collect telemetry data. &
  \component{Cubesat} shall \timing{always} \responseF{CollectData} &
  \textsc{Maintain Safe Space} &
  \cite{AuthorCubeSatFlightSoftwareFramework}  \\ \hline
28-B &
  in charge of providing the ground segment with telemetry data about the state and health of the spacecraft, therefore this service shall be able to automatically store telemetry data. &
  \component{Cubesat} shall \timing{always} \responseF{StoreData} &
  \textsc{Maintain Safe Space} &
  \cite{AuthorCubeSatFlightSoftwareFramework}  \\ \hline
28-C &
  in charge of providing the ground segment with telemetry data about the state and health of the spacecraft, therefore this service shall be able to automatically transmit telemetry data. &
  \conditionF{Whenever groundsegmentconnection \& !telemtrydata} \component{Cubesat} shall \timing{until telemtrydata} \responseF{ Transmit} &
  \textsc{Transmit} &
  \cite{AuthorCubeSatFlightSoftwareFramework} \\ \hline
\multicolumn{1}{|r|}{29} &
  At least one side shall be the pilot flying side. &
  \component{FGS} shall \timing{always} \responseF{PilotFlying \textless{}= 1} &
  \textsc{Semi-Autonomo{\-}us} &
  \cite{Author2023FormalMethodsCaseStudiesDO333} \\ \hline
\multicolumn{1}{|r|}{30} &
  At most one side shall be the pilot flying side. &
  \component{FGS} shall \timing{always} \responseF{PilotFlying \textgreater{}= 1} &
  \textsc{Semi-Autonomo{\-}us} &
  \cite{Author2023FormalMethodsCaseStudiesDO333}  \\ \hline
\multicolumn{1}{|r|}{31} &
  Pressing the Transfer Switch shall always change the pilot flying side. &
  \conditionF{Upon TransferSwitch} \component{FGS} shall \timing{immediately} \responseF{SwitchSides} &
  \textsc{Triggered Instant Reaction} &
  \cite{Author2023FormalMethodsCaseStudiesDO333} \\ \hline
\multicolumn{1}{|r|}{32} &
  The system shall start with the Primary Side as the pilot flying side. &
  \conditionF{Upon Startup} \component{FGS} shall \timing{at the next timepoint} \responseF{PrimarySide} &
  \textsc{Triggered Instant Reaction} &
  \cite{Author2023FormalMethodsCaseStudiesDO333} \\ \hline
\multicolumn{1}{|r|}{33} &
  The system shall not change the pilot flying side unless the Transfer Switch is pressed. &
  \component{FGS} shall \timing{until switch} \responseF{!SwitchSides} \qquad \qquad \qquad\qquad \texttt{+} \component{FGS} shall \timing{eventually} \responseF{switch} &
  \textsc{Wait} &
  \cite{Author2023FormalMethodsCaseStudiesDO333}  \\ \hline
\multicolumn{1}{|r|}{34} &
  Exceeding sensor limits shall latch an autopilot pullup when the pilot is not in control (not standby) and the system is supported without failures (not apfail). &
  \conditionF{Whenever Limits \& !Standby \& supported \& !apfail} \component{FSM} shall \timing{immediately} \responseF{Pullup} &
  \textsc{Instant Reaction} &
  \cite{katis2022realizability}\\ \hline
\multicolumn{1}{|r|}{35} &
  While flying, remain separated from an intruder aircraft by at least 250 ft horizontally or 50 ft vertically &
  \scope{In FlightMode} \component{AirCraft} shall \timing{always} \responseF{( horizontalIntruderDistance \textgreater 250 | verticalIntruderDistance \textgreater 50 )} &
  \textsc{Modal Maintain Safe Space} &
  \cite{agogino2024recommendations} \\ \hline
\multicolumn{1}{|r|}{36} &
  The probability that the aircraft leaves the taxiway, i.e., |cte| \textgreater 8 meters, shall be extremely low &
  \component{Aircraft} shall \probability{with probability \textless{}= 0.001} \timing{eventually} \responseF{absReal(cte) \textgreater 8} &
  \textsc{Probabilis{\-}tic Maintain Safe Space} &
  \cite{agogino2024recommendations}\\ \hline
\multicolumn{1}{|r|}{37} &
  The probability that the aircraft turns more than a prescribed degree (|he| $\leq$ 35°) shall be extremely low &
  \component{Aircraft} shall \probability{with probability  <= 0.002} \timing{eventually} \responseF{ absReal(he) <= 35} &
  \textsc{Probabilis{\-}tic Maintain safe Space} &
  \cite{agogino2024recommendations} \\ \hline
\multicolumn{1}{|r|}{38} &
  We also require that the rear propeller be always used, except in HC mode &
  \scope{If not in HCMode} \component{LPC} shall \timing{always} \responseF{ RearPropeller} &
  \textsc{Modal Maintain Safe Space} &
  \cite{Pressburger2023LiftPlusCruiseFRET} \\ \hline
\multicolumn{1}{|r|}{39} &
  If the vehicle is slowing down from the wing-borne mode (WB), the transition to semi-wing-borne (SWB) kicks in at an indicated airspeed of 90 knots (kias \textless{}= 90.0) &
  \scope{In Wbmode} \conditionF{whenever airspeed \textless{}= 90} \component{LPC} shall \timing{eventually} \responseF{SWBMode} &
  \textsc{Modal Delayed Reaction} &
  \cite{katis2022realizability}\\ \hline
\multicolumn{1}{|r|}{40} &
  whereas if the vehicle is speeding up from a SWB mode, the transition to WB mode occurs at kias \textgreater 100.0 knots &
  \scope{In SWBmode} \conditionF{whenever airspeed \textgreater 100} \component{LPC} shall \timing{eventually} \responseF{WBMode} &
  \textsc{Modal Delayed Reaction} &
  \cite{katis2022realizability} \\ \hline
\multicolumn{1}{|r|}{41} &
  The vehicle remains in the thrust-borne mode (TB) as long as kgs \textless{}= 20.0 knots and Hover Control (HC) mode is selected. &
  \scope{In HCmode} \conditionF{whenever TBMode \& Kgs \textless{}= 20} \component{LPC} shall \timing{always} \responseF{TBmode} &
  \textsc{Maintain Mode In Hierarchy} &
  \cite{Pressburger2023LiftPlusCruiseFRET} \\ \hline
\multicolumn{1}{|r|}{42-A} &
  during takeoff and landing, the aircraft motion is controlled by the lifting rotors only &
  \scope{In TakeoffMode} \component{LPC} shall \timing{always} \responseF{ LiftingRotors \& !FlightSurfaces} &
  \textsc{Modal Maintain Safe Space} &
  \cite{Pressburger2023LiftPlusCruiseFRET}\\ \hline
  \multicolumn{1}{|r|}{42-B} &
  during takeoff and landing, the aircraft motion is controlled by the lifting rotors only  &
  \scope{In LandingMode} \component{LPC} shall \timing{always} \responseF{ LiftingRotors \& !FlightSurfaces} &
  \textsc{Modal Maintain Safe Space}&
    \cite{Pressburger2023LiftPlusCruiseFRET}\\ \hline
\multicolumn{1}{|r|}{43} &
  On the other hand, during the higher speeds of the en-route phase, the wings provide lift, the rear propeller provides thrust, and the lifting rotors are inactive (wing-borne mode, WB) &
  \scope{In EnRoute} \component{LPC} shall \timing{always} \responseF{!LiftingRotors \& ThrustRearPropeller \& WingsLift} &
  \textsc{Modal Maintain Safe Space} &
  \cite{Pressburger2023LiftPlusCruiseFRET} \\ \hline
\multicolumn{1}{|r|}{44} &
 In a SLM survey, crew takes measurements at locations described in procedures, attempting to take the    measurement as close to the described point as possible &
  \component{Astrobee} shall \timing{eventually} \responseF{ SoundLocation \& SLMSurvey} &
  \textsc{Visit with Reaction} &
  \cite{Bualat2018Astrobee} \\ \hline
\multicolumn{1}{|r|}{45} &
 This type of data could be supplemented with denser, though shorter duration, measurements from a mobile platform. The Radiation Environment Monitor (REM) hardware developed at the University of Houston and NASA Johnson Space Center is an example of the sort of small, light-weight sensor that Astrobee could carry to create higher resolution maps of the ISS environment. &
  \component{Astrobee} shall \timing{eventually} \responseF{ RadiationLocation \& RadiationSurvey} &
 \textsc{Visit with Reaction} &
  \cite{Bualat2018Astrobee} \\ \hline
\multicolumn{1}{|r|}{46} &
  The SPHERES satellites, however, triangulate their position using infrared/ultrasonic beacons, preventing them from navigating outside the two-meter cube defined by the fixed beacon locations. &
  \conditionF{Whenever moving} \component{SPHERES} shall \timing{immediately} \responseF{ x\textless{}2 \& y\textless{}2 \& z\textless{}2} &
  \textsc{Stay-In-Perimeter} &
  \cite{IntBall22024RAIMag} \\ \hline
\multicolumn{1}{|r|}{47} &
  Like SPHERES, Int-Ball cannot operate without a direct line-of-sight to its markers. &
  \conditionF{whenever Operating} \component{IntBall} shall \timing{immediately} \responseF{LOS1 \& LOS2} &
  \textsc{Stay-In-Perimeter} &
  \cite{IntBall22024RAIMag} \\ \hline
\multicolumn{1}{|r|}{48} &
  The PerchCam is identical to the HazCam and it turns on to detect ISS handrails when Astrobee perches autonomously &
  \conditionF{Whenever Perched} \component{Astrobee} shall \timing{immediately} \responseF{PerchCam} &
  \textsc{Instant Reaction} &
  \cite{Bualat2018Astrobee} \\ \hline
\multicolumn{1}{|r|}{49} &
 the top-facing SpeedCam sensor package provides an independent over-speed cutoff function, estimating velocity using its own optical flow, infrared ranging, and IMU sensors &
  \conditionF{Whenever Moving} \component{Astrobee} shall \timing{always} \responseF{cutoff \textgreater currentspeed} &
  \textsc{Maintain Safe Space} &
  \cite{Bualat2018Astrobee} \\ \hline
\multicolumn{1}{|r|}{50} &
  After a sortie, Astrobee transfers large files through a hard-wired Ethernet connection with its dock &
  \conditionF{whenever ISSConnection \& Ethernet \& !LargeFile} \component{Astrobee} shall \timing{until LargeFile} \responseF{Transfer} &
  \textsc{Transmit} &
  \cite{Bualat2018Astrobee} \\ \hline
\multicolumn{1}{|r|}{51} &
  Once Astrobee grasps a handrail, it powers down its propulsion system. &
  \conditionF{Upon Perched} \component{Astrobee} shall \timing{at the next timepoint} \responseF{ !PropulsionSystem} &
  \textsc{Triggered Prompt Reaction}  &
  \cite{Bualat2018Astrobee} \\ \hline
\multicolumn{1}{|r|}{52} &
 Initially, Astrobee will use these components primarily to help crew understand its state and intentions (for example, by providing turn signals) &
  \conditionF{whenever Turning} \component{Astrobee} shall \timing{at the next timepoint} \responseF{Indicate} &
  \textsc{Prompt Reaction} &
  \cite{Bualat2018Astrobee} \\ \hline
\multicolumn{1}{|r|}{53} &
  When docking, Astrobee autonomously approaches its berth using visual servoing relative to fiducials mounted to the dock &
  \scope{In DockingMode} \component{Astrobee} shall \timing{eventually} \responseF{approachberth} &
   \textsc{Modal Reaction} &
  \cite{Bualat2018Astrobee} \\ \hline
\multicolumn{1}{|r|}{54} &
  When mating is complete, permanent magnets on the berth attract striker plates on the robot, providing a passive retention force &
  \conditionF{Upon MatingComplete} \component{DS} shall \timing{eventually} \responseF{StrikeMagnets} &
  \textsc{Triggered Delayed Reaction} &
  \cite{Bualat2018Astrobee} \\ \hline
\multicolumn{1}{|r|}{55} &
  To enable undocking, linear actuators within the berths pull the magnets away from the striker plates, allowing the propulsion system to easily overcome the reduced magnetic force &
  \scope{In UndockingMode} \component{DS} shall \timing{at the next timepoint} \responseF{LinearActuators} &
  \textsc{Modal Prompt Reaction} &
  \cite{Bualat2018Astrobee} \\ \hline
\multicolumn{1}{|r|}{56} &
  If multiple Astrobees are active, the Control Station displays the positions of all of the Astrobees so that the operators are aware of the other activities and can avoid collisions. &
  \conditionF{Whenever numberofAtrobees \textgreater{} 1} \component{CS} shall \timing{immediately} \responseF{DisplayALL} &
  \textsc{Instant Reaction} &
  \cite{Bualat2018Astrobee} \\ \hline
\multicolumn{1}{|r|}{57} &
Operators use the Plan Editor tab in the Control Station to construct and validate sequences of commands for        Astrobee (``.fplans''), that include waypoints and actions to perform at the waypoints. &
  \conditionF{whenever CommandReceived} \component{Astrobee} shall \timing{eventually} \responseF{ PerformCommand} &
  \textsc{Delayed Reaction} &
  \cite{Bualat2018Astrobee} \\ \hline
58-A &
  Astrobee can lose signal, when signal is lost Astrobee enters LOSMode &
  \conditionF{whenever LostSignal} \component{Astrobee} shall \timing{immediately} \responseF{LOSMode} &
  \textsc{Instant Reaction} &
  \cite{Bualat2018Astrobee} \\ \hline
58-B &
 During loss-of-signal (LOS) with the ground, Astrobee continues to hold its position while recording and storing video on its internal file system. &
  \scope{In LOSMode} \component{Astrobee} shall \timing{always} \responseF{Hold \& WorkInternally} &
  \textsc{Modal Maintain Safe Space} &
  \cite{Bualat2018Astrobee} \\ \hline
58-C &
  Once ground signal has been reacquired &
  \conditionF{Whenever !ISSConnection \& !Groundsignal} \component{Astrobee} shall \timing{until ISSConnection \& Groundsignal} \responseF{reconnect} &
  \textsc{Reconnect} &
  \cite{Bualat2018Astrobee} \\ \hline
58-D &
  Once ground signal has been reacquired, Astrobee resumes downlinking the live video stream to the Control Station. &
  \conditionF{Whenever ISSConnection \& Groundsignal \& !Stream} \component{Astrobee} shall \timing{until Stream} \responseF{downlink} &
  \textsc{Transmit} &
  \cite{Bualat2018Astrobee} \\ \hline
\multicolumn{1}{|r|}{59-A} &
  Astrobee is programmed to stop when it detects an obstacle &
  \conditionF{Upon ObstacleDetected} \component{Astrobee} shall \timing{immediately} \responseF{Stop} &
  \textsc{Triggered Instant Reaction}&
  \cite{Bualat2018Astrobee} \\ \hline
\multicolumn{1}{|r|}{59-B} &
  we are considering using Astrobee’s lights and/or speaker to signal when it enters a hatchway &
  \conditionF{Whenever EntersHatchway} \component{Astrobee} shall \timing{immediately} \responseF{ EntranceAlarm} &
  \textsc{Instant Reaction} &
  \cite{Bualat2018Astrobee} \\ \hline
60-A &
  White ``Vid'' LEDs indicate that cameras are in use &
  \scope{In VideoRecordingMode} \component{Astrobee} shall \timing{always} \responseF{VidLED} &
  \textsc{Modal Maintain Safe Space} &
  \cite{Bualat2018Astrobee} \\ \hline
60-B &
  A blue ``Aud'' light tells the crew that the microphone is on &
  \scope{In AudioRecording} \component{Astrobee} shall \timing{always} \responseF{BlueAudLED} &
  \textsc{Modal Maintain Safe Space} &
  \cite{Bualat2018Astrobee} \\ \hline
60-C &
  ``Live'' LEDs indicate that cameras are streaming &
  \scope{In StreamingMode} \component{Astrobee} shall \timing{always} \responseF{LiveLED} &
  \textsc{Modal Maintain Safe Space} &
  \cite{Bualat2018Astrobee} \\ \hline
\multicolumn{1}{|r|}{61-A} &
 The Control Station warns operators when they create plans that translate through a KOZ. &
  \conditionF{Whenever KOZPlan} \component{ControlStation} shall \timing{at the next timepoint} \responseF{Warn} &
  \textsc{Prompt Reaction}
   &
  \cite{Bualat2018Astrobee} \\ \hline
  \multicolumn{1}{|r|}{61-B} &
 The Control Station prevents operators from sending plans that translate through a KOZ to Astrobee until the violating segments are modified &
 \component{ControlStation} shall \timing{until !KOZPlan} \responseF{!SendPlan} \qquad\qquad \texttt{+} \component{ControlStation} shall \timing{eventually}  \responseF{!KOZPlan}  &
  \textsc{Wait}
   &
  \cite{Bualat2018Astrobee} \\ \hline
\multicolumn{1}{|r|}{62} &
  As a final safeguard, Astrobee itself has an internal list of KOZs that it checks before moving &
  \conditionF{Whenever moving} \component{Astrobee} shall \timing{immediately} \responseF{!KOZ1 \& !KOZ2} &
  \textsc{Keep-Out-Zone} &
  \cite{Bualat2018Astrobee} \\ \hline
63-A &
  The robot can periodically update multi-sensor 3D maps of the vehicle. air quality tracking can all help flight controllers understand system status &
  \conditionF{Whenever TimeForAir} \component{Astrobee} shall \timing{eventually} \responseF{AirSurvey} &
  \textsc{Delayed Reaction} &
  \cite{Bualat2018Astrobee} \\ \hline
63-B &
  The robot can periodically update multi-sensor 3D maps of the vehicle. RFID quality tracking can all help flight controllers understand system status &
  \conditionF{Whenever TimeForRFID} \component{Astrobee} shall \timing{eventually} \responseF{RFIDSurvey} &
  \textsc{Delayed Reaction} &
  \cite{Bualat2018Astrobee} \\ \hline
63-C &
   The robot can periodically update multi-sensor 3D maps of the vehicle. Visual Imaging tracking can all help flight controllers understand system status &
   \conditionF{Whenever TimeForVisualImaging} \component{Astrobee} shall \timing{eventually} \responseF{VisualImagingSurvey} &
  \textsc{Delayed Reaction} &
  \cite{Bualat2018Astrobee} \\ \hline
63-D &
  The robot can periodically update multi-sensor 3D maps of the vehicle. Thermal imaging tracking can all help flight controllers understand system status &
  \conditionF{Whenever TimeForThermalImaging} \component{Astrobee} shall \timing{eventually} \responseF{ ThermalImagingSurvey} &
  \textsc{Delayed Reaction} &
  \cite{Bualat2018Astrobee} \\ \hline
\multicolumn{1}{|r|}{64} &
  Automated change detection and trending. Once a baseline sensor map is available, changes at the next update can indicate developing problems at an early stage &
  \conditionF{Whenever SurveyDone} \component{Astrobee} shall \timing{eventually} \responseF{CompareMaps} &
  \textsc{Delayed Reaction} &
  \cite{Bualat2018Astrobee} \\ \hline
\multicolumn{1}{|r|}{65} &
  Localizing problems. For example, if a leak produces a whistling sound, acoustic or ultrasonic sensors onboard the robot can be used to pinpoint its location. &
  \conditionF{Whenever AnomalyDetected} \component{Astrobee} shall \timing{eventually} \responseF{PinpointProblem} &
  \textsc{Delayed Reaction} &
  \cite{Bualat2018Astrobee} \\ \hline
\multicolumn{1}{|r|}{66} &
  When flight controllers have a question about something, they can use the robot to get an updated view, filling a role currently played by crew on ISS &
  \conditionF{Whenever SpotCheck} \component{Astrobee} shall \timing{eventually} \responseF{UpdateMap} &
  \textsc{Delayed Reaction} &
  \cite{Bualat2018Astrobee} \\ \hline
\multicolumn{1}{|r|}{67} &
  The first is the observing and planning phase for acquiring motion information of the target satellite and planning when and where the robot will grasp the target satellite &
  \conditionF{Upon FirstPhase} \component{ServicingSatellite} shall \timing{eventually} \responseF{AcquireMotionInformation \& Planning} &
  \textsc{Phases} &
  \cite{Pham2014RoboticsOnOrbitServicingReview} \\ \hline
\multicolumn{1}{|r|}{68} &
  The second phase is to control the robot to move toward the planned grasping location to make the robot ready for the capturing of the target. &
 \conditionF{Upon AcquireMotionInformation \& Planning} \component{ServicingSatellite} shall \timing{at the next timepoint} \responseF{SecondPhase} \qquad \qquad \qquad \texttt{+} \conditionF{Upon SecondPhase} \component{ServicingSatellite} shall \timing{eventually} \responseF{MoveToPosition} &
  \textsc{Phases} &
  \cite{Pham2014RoboticsOnOrbitServicingReview} \\ \hline
\multicolumn{1}{|r|}{69} &
  The third phase is the capture (physical interception) phase in which the manipulator physically captures the target satellite &
  \conditionF{Upon MoveToPosition} \component{ServicingSatellite} shall \timing{at the next timepoint} \responseF{ThirdPhase} \qquad \qquad \qquad \qquad \texttt{+} \conditionF{Upon ThirdPhase} \component{RobotManipulator} shall \timing{eventually} \responseF{PhysicalCapture} &
  \textsc{Phases} &
  \cite{Pham2014RoboticsOnOrbitServicingReview} \\ \hline
\multicolumn{1}{|r|}{70} &
  The fourth phase is the post-capture phase in which captured target satellite is stabilized along with the servicing system &
  \conditionF{Upon PhysicalCapture} \component{ServicingSatellite} shall \timing{at the next timepoint} \responseF{FourthPhase} \qquad \qquad \qquad \qquad \qquad\texttt{+} \conditionF{Upon FourthPhase} \component{ServicingSatellite} shall \timing{eventually} \responseF{Stabilization} &
  \textsc{Phases} &
  \cite{Pham2014RoboticsOnOrbitServicingReview} \\ \hline
\multicolumn{1}{|r|}{71} &
 The maximum rotation speed is restricted within 23,100 rpm to ensure the crew’s safety &
  \component{IntBall2} shall \timing{always} \responseF{RPM \textless{}= 23100} &
  \textsc{Maintain Safe Space} &
 \cite{IntBall22024RAIMag} \\ \hline
\multicolumn{1}{|r|}{72} &
 Once the variances of the derivatives of acceleration and angular velocity from the IMU exceed pre-defined upper thresholds, the status shifts to collision mode. &
  \conditionF{If (VelocityVariancesc \textgreater UpperVelocityThreshold) \& (AccelerationVariances \textgreater   UpperAccelerationThreshold)} \component{IntBall2} shall \timing{immediately} \responseF{CollisionMode} &
  \textsc{Triggered Instant reaction} &
 \cite{IntBall22024RAIMag} \\ \hline
\multicolumn{1}{|r|}{73} &
 If the variances immediately decrease below lower thresholds, the impact cause is presumed to be an impulsive external force, and the Int-Ball2 tries to maintain its current pose &
  \scope{In CollisionMode} \conditionF{if (VelocityVariances \textless LowerVelocityThreshold) \& (AccelerationVariances \textless LowerAccelerationThreshold)} \component{IntBall2} shall \timing{immediately} \responseF{MaintainCurrentPose} &
  \textsc{Modal Triggered Instant reaction} &
 \cite{IntBall22024RAIMag} \\ \hline
\multicolumn{1}{|r|}{74} &
 Otherwise, it is assumed to be held by the astronaut’s hands, and maneuver control is turned off &
  \scope{In CollisonMode} \conditionF{if !(VelocityVariances \textless LowerVelocityThreshold) | !(AccelerationVariances \textless LowerAccelerationThreshold)} \component{IntBall2} shall \timing{immediately} \responseF{ManeuverControl=0 \& AstronautControl} &
  \textsc{Modal Triggered Instant Reaction} &
 \cite{IntBall22024RAIMag} \\ \hline
\multicolumn{1}{|r|}{75} &
  After the astronaut releases the robot, the control for maintaining the pose at the released point is restarted if the variance falls below the lower threshold &
  \conditionF{If AstronautControl=0 \& VarianceThreshold \textless VarianceThreshold} \component{IntBall2} shall \timing{immediately} \responseF{MaintainCurrentPose} &
  \textsc{Triggered Instant Reaction} &
 \cite{IntBall22024RAIMag} \\ \hline
\multicolumn{1}{|r|}{76} &
 Additionally, when the navigation camera is blocked by crew interference or positioned too close to a wall so that the feature points for vSLAM cannot be detected, the navigation subsystem shifts to inertial navigation that uses the IMU without relying on the vSLAM output. &
  \conditionF{Whenever vSLAMUnavailable} \component{IntBall2} shall \timing{at the next timepoint} \responseF{NavigatewithIMU \& NavigatevSLAM=0} &
  \textsc{Prompt Reaction} &
 \cite{IntBall22024RAIMag} \\ \hline
77-A &
 However, if the vSLAM output remains unavailable for an extended period, the robot rotates in place until the feature points detected in the current view align with those in the stored map &
  \conditionF{whenever vSLAMUnavailable} \component{IntBall2} shall \timing{until FeaturePointDetected} \responseF{RotateProtocol} \qquad \qquad \qquad \qquad \texttt{+} \conditionF{whenever vSLAMUnavailable} \component{IntBall2} shall \timing{eventually} \responseF{FeaturePointDetected} &
  \textsc{Conditional Wait} &
 \cite{IntBall22024RAIMag} \\ \hline
77-B &
  if the vSLAM output remains unavailable for an extended period &
  \conditionF{Whenever vSLAMOutput=0 \& TimePassed \textless{}= ExtendedPeriod} \component{IntBall2} shall \timing{at the next timepoint} \responseF{vSLAMUnavailable} &
  \textsc{Prompt Reaction} &
  \cite{IntBall22024RAIMag} \\ \hline
\multicolumn{1}{|r|}{78} &
Furthermore, when the Int-Ball2 automatically detects that the remaining battery power is low, it returns to the DS for recharging &
  \conditionF{Whenever IntBall2Power \textless{}= SafeBattery} \component{IntBall2} shall \timing{at the next timepoint} \responseF{RechargeMode} &
 \textsc{Prompt Reaction} &
  \cite{IntBall22024RAIMag} \\ \hline
\end{longtable}
\end{document}